\documentclass[conference,compsoc]{IEEEtran}
\IEEEoverridecommandlockouts
\usepackage{cite}
\usepackage{amsmath,amssymb,amsfonts}
\usepackage{algorithmic}
\usepackage{graphicx}
\usepackage{textcomp}
\usepackage{xcolor}
\def\BibTeX{{\rm B\kern-.05em{\sc i\kern-.025em b}\kern-.08em
    T\kern-.1667em\lower.7ex\hbox{E}\kern-.125emX}}
\usepackage{fontawesome}
\usepackage{booktabs,tabularx}
\usepackage{enumitem}
\usepackage{adjustbox}
\usepackage{array}
\usepackage{booktabs}
\usepackage{multirow}
 \usepackage{graphicx}
 \usepackage[hyphens]{url}
 \usepackage{epsfig,endnotes}
\usepackage{xcolor}
\usepackage{color}
\usepackage{colortbl}
\usepackage{graphicx}
\usepackage{multirow}
\usepackage{listings}
\usepackage[ruled,linesnumbered,lined,boxed]{algorithm2e}
\usepackage{tabularx}
\usepackage{tikz}
\usepackage[normalem]{ulem}
\useunder{\uline}{\ul}{}
\usepackage{ragged2e}  
\usepackage{balance}
\usepackage{xspace}
\usepackage{paralist}
\usepackage{pifont}
\usepackage{wasysym}
\usepackage{amsmath}
\usepackage[binary-units=true]{siunitx}
\usepackage{amsthm}
\usepackage{amssymb}
\usepackage{url}
\usepackage{booktabs}
\usepackage{dcolumn}
\usepackage{MnSymbol}
\usepackage{comment}
\usepackage{multirow}
\usepackage{makecell}
\usepackage{tcolorbox}

\definecolor{darkcerulean}{rgb}{0.03, 0.27, 0.49}
\definecolor{darkblue}{rgb}{0.0, 0.28, 0.67}
\definecolor{Gray}{gray}{0.85}
\definecolor{Gray2}{gray}{0.95}
\newenvironment{takeaway}[1]{%
  \begin{tcolorbox}[colback=Gray2,colframe=darkcerulean,title=\textup{\textbf{#1}}]
}{%
  \end{tcolorbox}
}

\newcolumntype{a}{>{\columncolor{Gray}}c}

\newcolumntype{R}[2]{%
    >{\adjustbox{angle=#1,lap=\width-(#2)}\bgroup}%
    l%
    <{\egroup}%
}

\definecolor{darkgreen}{rgb}{0.0, 0.5, 0.0}
\definecolor{babyblueeyes}{rgb}{0.63, 0.79, 0.95}
\newcommand{\cmark}{\textcolor{darkgreen}{\ding{51}}}

\newcommand{\xmark}{\textcolor{red}{\ding{55}}}
\newcommand{\xmarkgood}{\textcolor{darkgreen}{\ding{55}}}
\newcommand{\nmarkr}{\textcolor{red}{\ding{108}}}
\newcommand{\nmarkg}{\textcolor{darkgreen}{\ding{108}}}
\newcommand{\nmarko}{\textcolor{orange}{\ding{108}}}
\newcommand{\mah}{\textcolor{red}}

\newcommand*\circled[1]{\tikz[baseline=(char.base)]{
            \node[shape=circle,fill,inner sep=1pt] (char) {\textcolor{white}{#1}};}}

\newcommand{\cflog}{{\ensuremath{\sf{\textsc{CF$_{Log}$}}}}\xspace}

\newcommand{\rep}{{\ensuremath{\sf{\textsc{CF$_{Report}$}}}}\xspace}

\newcommand{\vrf}{{\ensuremath{\sf{\mathcal Vrf}}}\xspace}
\newcommand{\prv}{{\ensuremath{\sf{\mathcal Prv}}}\xspace}

\newcommand{\advs}{{\ensuremath{\sf{\mathcal Adv}}-s}\xspace}

\newcommand{\halftriangle}{
\begin{tikzpicture}[scale=0.35]
    \draw (0,0) -- (.5,0) -- (0.25,.5) -- cycle;
    \fill[black] (0,0) -- (0.25,0) -- (0.25,.5) -- cycle;
\end{tikzpicture}
}

\definecolor{altgreen}{RGB}{0,150,0}

\newlist{minimalitemize}{itemize}{1}
\setlist[minimalitemize]{
    label=$\bullet$,
    align=left,
    leftmargin=*,
    nosep,
}

\newlist{minimalenum}{enumerate}{1}
\setlist[minimalenum]{
  label=\arabic*),
  align=left,
  leftmargin=*,
  nosep,
}

\newlist{questionlist}{enumerate}{1}
\setlist[questionlist]{
  label=\textbf{[Q\arabic*}],
  align=left,
  leftmargin=*,
  nosep,
}

\begin{document}

\title{SoK: Runtime Integrity}

\author{
\IEEEauthorblockN{Mahmoud Ammar}
\IEEEauthorblockA{Independent Researcher\\
mail@mahmoud-ammar.org}
\and
\IEEEauthorblockN{Adam Caulfield}
\IEEEauthorblockA{Rochester Institute of Technology\\
ac7717@rit.edu}
\and
\IEEEauthorblockN{Ivan De Oliveira Nunes}
\IEEEauthorblockA{Rochester Institute of Technology\\
ivanoliv@mail.rit.edu}
}

\maketitle

\begin{abstract}
This paper provides a systematic exploration of Control Flow Integrity (CFI) and Control Flow Attestation (CFA) mechanisms, examining their differences and relationships. It addresses crucial questions about the goals, assumptions, features, and design spaces of CFI and CFA, including their potential coexistence on the same platform. Through a comprehensive review of existing defenses, this paper positions CFI and CFA within the broader landscape of runtime defenses, critically evaluating their strengths, limitations, and trade-offs. The findings emphasize the importance of further research to bridge the gaps in CFI and CFA and thus advance the field of runtime defenses. 
\end{abstract}


\begin{figure}[!b]
\begin{minipage}{\columnwidth}
  \rule{\linewidth}{0.4pt}
  \noindent\small{\textbf{To appear: \textit{IEEE Security \& Privacy 2025 (S\&P 2025).} \\
  Title: \textit{SoK: Integrity, Attestation, and Auditing of Program Execution}}}.
\end{minipage}
\end{figure}

\section{Introduction}
\label{sec:introduction}

Unsafe programming languages like C and C++ are still prevalent, especially for lower-level system development~\cite{pl-index}. Memory safety bugs, such as buffer overflows, are prominent enablers of attacks on programs written in such languages. 
Attacks that modify/inject code can be (to some extent) mitigated by existing defenses. Among them, Data Execution Prevention (DEP) and Write-Xor-eXecute (W$\oplus$X)~\cite{dep} policies can prevent user-space code injection attempts at runtime. Secure boot can locally enforce boot-time code integrity (including the integrity of privileged software, e.g., stage 1 and 2 boot-loaders and kernel)\cite{lohr2010patterns, arbaugh1997secure}. Static (i.e., boot-time or load-time) Remote Attestation (RA) can further convince a remote party of the integrity of the booted code chain~\cite{ima-lwn,android-attest}.

On the other hand, code-reuse attacks~\cite{codereuse-attacks} (exemplified by Return Oriented Programming -- ROP~\cite{rop} -- and Jump Oriented Programming -- JOP ~\cite{jop}) can pose significant threats without modifying the installed code, even in the presence of existing defenses. They instead exploit memory corruption vulnerabilities to trigger out-of-order execution of sub-sequences of instructions (known as gadgets) within a program. This can result in unintended behavior even when code modifications are prevented.
As illustrated in Figure~\ref{fig:code-reuse-taxonomy}, code-reuse attacks can be broadly classified into two categories: control flow hijacking and data-only attacks. The former directly corrupts memory storing code pointers, e.g., return addresses~\cite{rop} and function pointers~\cite{jop} during execution. The latter changes control flow related data, e.g., loop/conditional variables or counters, without causing control flow transfers that do not exist in the Control Flow Graph (CFG) of the target program~\cite{non-control-data,dop}.

Much attention has been devoted to code-reuse attack mitigations due to their popularity and effectiveness~\cite{payer2017control} with several protection and detection mechanisms proposed in the past few decades~\cite{eternal-war}. Most notably, Control Flow Integrity (CFI) mechanisms for both forward and backward edge protection have been widely recognized as key mitigations~\cite{original-cfi,cfi-survey-1}. Ergo, recent years have seen efforts to adopt both academic and industry proposals, each with their own sets of trade-offs~\cite{cfi-survey-1,cfi-survey-2,confirm}. Nevertheless, only a few of these proposals, e.g., LLVM CFI~\cite{llvm-cfi}, have become available in production compilers, despite known limitations in terms of granularity and compatibility~\cite{cfi-bending,confirm}.

On the hardware side, both ARM and Intel have equipped their latest-generation architectures with new extensions to assist control flow attack mitigations. Examples include Pointer Authentication (PA), Memory Tagging Extension (MTE), and Branch Target Identification (BTI) features from ARM~\cite{pac-bti-mte}, and the Control Flow Enforcement Technology (CET) from Intel~\cite{intel-cet}. While various contemporary CFI approaches leverage these extensions in their designs~\cite{fineibt,apple-pac,google-mte,pac-it-up,pal,pacstack,serra2022pac,color-my-world}, 
gaps still persist~\cite{pacman,azad2021examining,ios-pac}.

In a parallel line of efforts, Control Flow Attestation (CFA)~\cite{c-flat,tinycfa,lofat,atrium,litehax,oat,blast,ari,acfa,traces,isc-flat} has been proposed to enable remote Verifier(s) (\vrf) to ascertain the execution integrity (including the absence of control flow attacks/violations) of an operation of interest performed by a remote device (called a prover or \prv).
In its ideal form, CFA generates an authenticated log containing all dynamically defined control transfers occurring during the execution of an attested operation of interest\footnote{Note that some CFA variations aim at enabling continuous verification of all control flow transfers on \prv, rather than focusing on individual operations of interest. For more details, see Section~\ref{sec:definitions}.}. 
Nonetheless, similar to CFI, coarser-grained CFA approaches are also possible~\cite{ari}, establishing trade-offs between the completeness of CFA evidence and performance, especially when overheads related to storage and transmission of said evidence to \vrf are a concern (e.g., when \prv is a resource-constrained embedded platform).

Notably, not all CFA techniques guarantee that CFA evidence is received by \vrf.  While this is sufficient for attestation, wherein a \vrf would not trust responses/values received from \prv unless accompanied by CFA evidence, it does not support secure runtime auditing~\cite{acfa,traces}.
The latter aims to ensure that CFA evidence always reaches \vrf, even if \prv is compromised, allowing for attack root cause analysis and appropriate remediation.


CFI and CFA goals can be viewed as runtime analogs of boot-time code integrity guarantees offered by secure boot vs. static RA. While CFI enables \textit{in loco} detection of control flow violations (typically triggering exceptions when detected), CFA provides remotely verifiable (unforgeable) evidence of the control flow path followed by an operation of interest executed by a \prv device, thus enabling control flow path analysis by a remote \vrf.

\subsection{Motivation \& Intended Contributions}
Although CFI and CFA approaches exist due to the common threat of control flow attacks, their different goals, designs, and capabilities are not yet systematically discussed in the literature.
Naturally, the current lack of systematization prompts questions such as:






\begin{questionlist}
    \item How do CFA and CFI goals differ?
    \item What are the assumptions, features, and design spaces of CFI vs. CFA, as well as their similarities and differences?
    \item What makes CFA different from remotely attesting adherence to a CFI policy? Could CFA uncover attacks that CFI would not (and vice-versa)?
    \item Could CFI and CFA coexist on the same platform?
\end{questionlist}


Additionally, there is often confusion surrounding the terminology in the context of control flow-related mechanisms (e.g., prevention vs. local detection vs. remote detection; runtime attestation vs. runtime auditing; fine-grained vs. coarse-grained approaches; etc.) and their relationship to memory safety and compartmentalization defenses. This ambiguity makes it challenging to precisely understand the guarantees provided by each approach. Therefore, it becomes crucial to delve into such nuances to clearly grasp the benefits of each approach and their roles within the broader landscape of runtime software defenses.

In this paper, we explore the relationships and differences between CFI and CFA by systematically examining the fundamental goals and trade-offs associated with both approaches. Towards this goal, we present a systematic review of existing runtime defenses to provide context and position CFI and CFA within the broader landscape of execution integrity defenses.
Subsequently, we classify recent work in CFI and CFA according to design choices, weighing their advantages and disadvantages and aiming to grasp a better understanding of existing limitations.
Finally, we discuss missing links between CFI and CFA and future research avenues.

\subsection{Literature Selection Criteria}

The selection criteria for the inclusion of academic or industrial proposals in our systematization are as follows:

\begin{minimalitemize}
\item We aim to include all available literature on CFA due to the manageable number of existing proposals (except unintended oversights).
\item Given the extensive volume of CFI proposals, we use the following criteria for selection within the past 10 years:
\begin{compactitem}
    \item Papers published in prestigious security-focused conferences such as USENIX Security, IEEE S\&P, CCS, and NDSS.
    \item Papers with more than 100 citations, indicating their broad influence in subsequent work.
    \item Papers or proposals adopted in mainstream compilers or hardware architectures.
\end{compactitem}
\end{minimalitemize}

\subsection{Scope \& Related Systematizations}\label{sec:scope}

Memory safety~\cite{memsafe1,memsafe2,memsafe3,memsafe4} approaches aim to eliminate or reduce vulnerabilities that could lead to control/data flow attacks and data corruption during software development, i.e., before deployment. These typically work in two ways. First, memory safety can be a built-in security feature of programming languages such as Go and Rust. Rust~\cite{rust}, for instance, utilizes static compile-time analysis to optimize safety checks and memory management decisions, such as bounds check elimination, while incorporating mechanisms (e.g., value ownership and borrowing) to ensure temporal safety.
Second, memory safety can be obtained as memory-safe dialects of memory-safe programming languages. An example of this is Checked C~\cite{checkedc}, which augments C with spatial memory safety checks introduced at compilation time and/or runtime. This involves refining the C type system with safe pointer and array types with stricter usage models. Even so, this approach provides partial protection and presents compatibility challenges with legacy software.

A third category of techniques focuses on fortifying and isolating code through runtime checks~\cite{nyman2020toward}. These techniques include compartmentalization~\cite{cheri}, software fault isolation~\cite{sfi}, memory layout randomization~\cite{aslr}, as well as CFI. While the primary goal of this class is to detect and isolate runtime violations, some literature still categorizes these methods as memory safety techniques. There is, however, no clear consensus on whether the term “memory safety” should be limited to the first two categories or expanded to include this third one (and possibly others). This lack of agreement has led to confusion over the scope of the term~\cite{memsafe1}.


Terminology aside, our work focuses on systematizing and discussing the relationship between runtime integrity enforcement~\cite{original-cfi} and runtime attestation~\cite{c-flat} methods used after software deployment (hence ``runtime''). This is complemented by existing systematizations focused, for instance, on memory safety or compartmentalization. Related to our work, Szekeres et al.~\cite{eternal-war} provide a general model of memory corruption attacks, which serves as a foundation for identifying the different policies that can prevent such attacks. Song et al.~\cite{sok-sanitizing} offer a systematic overview of sanitizers, emphasizing their role in uncovering security vulnerabilities. Larsen et al.~\cite{sok-sw-diversity} present a comprehensive and unified overview of software diversification approaches, highlighting their inherent trade-offs. Burow et al.~\cite{sok-shadow-stacks} conduct a thorough evaluation of the design space of shadow stacks, considering performance, compatibility, and security aspects. Contrary to the aforementioned efforts, this SoK focuses on shedding light and evaluating recent CFI and CFA methods as well as their relationship and differences.

\section{A Lightning Tour}
\label{sec:tour}

This section reviews code-reuse attacks and existing defenses, highlighting the role of CFI and CFA in this landscape.


\begin{figure}[t]
	\centering
	\includegraphics[width=0.95\columnwidth]{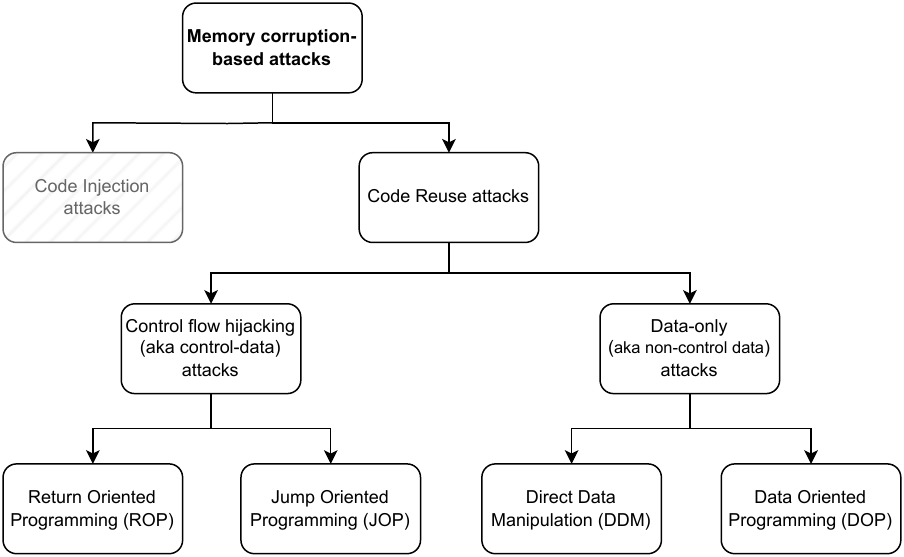}
    \vspace{-0.5em}
	\caption{Classes of memory corruption-based attacks to software integrity.}
	\label{fig:code-reuse-taxonomy}
\end{figure}

\subsection{Code Reuse Attack Background}
\label{subsec:attacks}

\begin{figure}[t]
	\centering
	\includegraphics[width=\columnwidth]{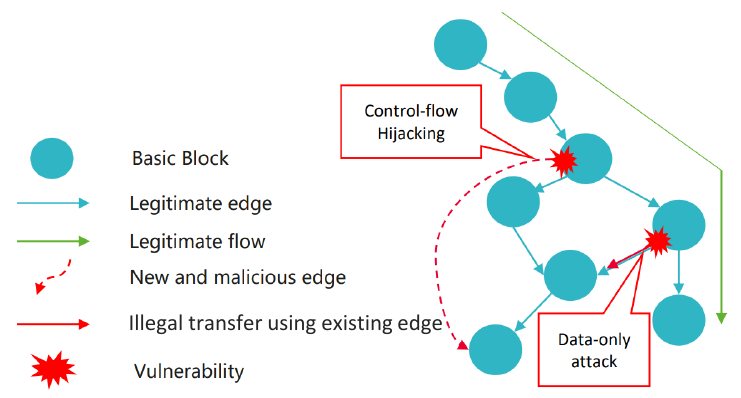}
    \vspace{-1.5em}
	\caption{Control flow hijacking vs. Data-only attacks on a CFG.}
	\label{fig:CFHvsDOA}
    \vspace{-1.5em}
\end{figure}

Figure~\ref{fig:code-reuse-taxonomy} illustrates the general classes of memory corruption-based attacks. Code-reuse attacks are further classified into control flow hijacking and data-only attacks. At a high level, their difference lies in the former performing control flow transfers that do not exist in the legitimate CFG of the target program and the latter causing unintended transfers via edges that exist in the CFG. The two cases are depicted in Figure~\ref{fig:CFHvsDOA}. Return Oriented Programming (ROP)~\cite{rop} and Jump Oriented Programming (JOP)~\cite{jop} are the two main categories of control flow hijacking attacks.
Both ROP and JOP stitch out-of-order sub-sequences of instructions, so-called \textit{gadgets}, to modify the control flow path of the target program to perform a malicious action. As their names indicate,
ROP corrupts backward edges, targeting gadgets
that end with \texttt{return} instructions. JOP corrupts forward edges, targeting gadgets that end with indirect \texttt{jump} or \texttt{call} instructions. 

Data-only attacks can be classified into Direct Data Manipulation (DDM) and Data Oriented Programming (DOP) based on the type of non-control data being manipulated~\cite{cheng2021exploitation}.
DDM attacks can be as simple as illegally modifying the value of a variable~\cite{non-control-data}.
DOP attacks~\cite{dop}, on the other hand, aim to perform expressive  (often Turing-complete) computations
by chaining carefully selected DOP gadgets, ensuring that the gadget chain forms a valid path within the CFG. 
This is typically achieved by corrupting non-control data, such as variables that define paths in conditional statements and loop counters.

\begin{figure*}[!t]
	\centering
	\includegraphics[width=.98\textwidth]{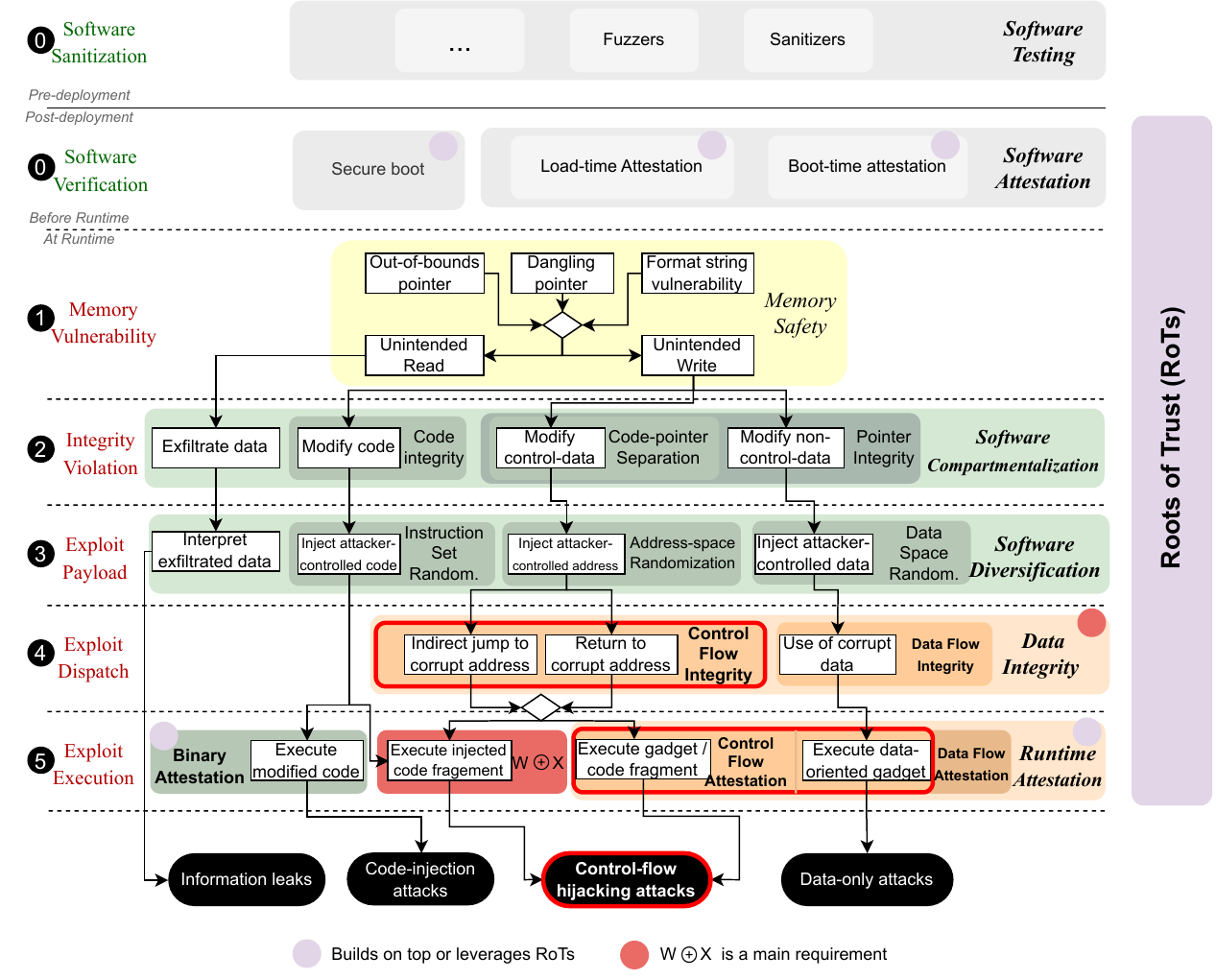}
    \vspace{-1em}
	\caption{A high-level overview of defenses against memory corruption-based attacks with a focus on runtime defenses (expanded based on~\cite{eternal-war} and~\cite{nyman2020toward}).
    }  
    \vspace{-1.5em}
	\label{fig:defenses}
\end{figure*}

\subsection{Runtime Defenses}
\label{subsec:defenses}

Figure~\ref{fig:defenses} illustrates the relationship between memory vulnerabilities, runtime exploits, and associated defenses.
The primary categories of runtime defenses against memory corruption-based attacks are illustrated in (\circled{2} - \circled{5}), which have been adapted from~\cite{eternal-war} and~\cite{nyman2020toward} respectively. Software testing tools, such as sanitizers~\cite{sok-sanitizing} and fuzzers~\cite{godefroid2020fuzzing}, act as a front-line defense in the pre-deployment phase, where the main goal is to find as many vulnerabilities as possible and fix them.  Boot- and Load-time software verification mechanisms, such as Secure Boot~\cite{boot}, Measured Boot~\cite{dice-boot}, and Load-time attestation (e.g., the Linux Integrity Measurement Architecture (IMA) for user-space software~\cite{ima}), are deployed as a primary shield in the post-deployment phase to prevent booting/loading of non-authentic software.  
However, the presence of memory corruption vulnerabilities at runtime remains a concern after this stage.
Therefore, several runtime defenses have been proposed, each targeting specific steps of the attack process.
Considering the five distinct attack steps (\circled{1} - \circled{5}) outlined in the general model of memory corruption attacks from~\cite{eternal-war}, Figure~\ref{fig:defenses} illustrates which class of defenses can counter each type of exploit and at which step. 
As information leaks are not integrity violations, they are not considered in Figure~\ref{fig:code-reuse-taxonomy}. In the following, we summarize the individual attack steps and relevant defenses: 

    \circled{1} \textbf{Memory Vulnerability: } Finding and exploiting a memory corruption vulnerability is an essential requirement for any of the runtime attacks considered in Figure~\ref{fig:defenses}. Illegal access to a memory address, whether to read, write, or both, depends on the particular vulnerability. We note that vulnerabilities that enable read-only access are (by themselves) not sufficient to corrupt the execution integrity of the target program. 

    \circled{2} \textbf{Integrity Violation: } Exploiting vulnerabilities that grant illegal write access enables adversaries to tamper with the various aspects of a program, including the (i) program's code (instructions in memory), (ii) control data (e.g., \texttt{return} addresses and function pointers), and (iii) non-control data (e.g., data variables and pointers). Isolation and compartmentalization mechanisms play a crucial role in enforcing access control permissions to mitigate integrity violations. These mechanisms restrict the targets that adversaries can access, often thwarting attacks at an early stage or preventing their spread to the rest of the system.  
    For instance, access control mechanisms like the AArch64 (Un)Privileged Execution Never feature~\cite{holdings2022armv8} make it significantly harder to directly corrupt program code~\cite{norax}. Code Pointer Integrity (CPI)\cite{cps} is a security mechanism that safeguards all code pointers and data pointers pointing to code by storing them in an isolated memory area. Code Pointer Separation (CPS)\cite{cps}, a variant of CPI, isolates only code pointers while leaving the protection of data pointers to other measures for performance reasons.
    Software Fault Isolation  (SFI)~\cite{sfi}, memory tagging~\cite{memory-tagging}, and capability-based architectures (exemplified by CHERI~\cite{cheri}) operate at various granularity levels to isolate larger software components into distinct protection domains. These mechanisms limit the consequences of attacks that exploit memory vulnerabilities by confining them within specific compartments.
    
    \circled{3} \textbf{Exploit Payload: } If previous defenses are bypassed, the adversary can inject payloads to manipulate the data and control flows of the target program. In general, the payload injection process requires knowledge of the program's memory layout. In response, software diversification aims to impede the crafting of reusable exploits by introducing uncertainty through randomization. For instance, Address Space Layout Randomization (ASLR)~\cite{aslr} and Instruction Set Randomization (ISR)~\cite{isr} are lightweight defenses that randomize memory layouts, making payload injection more challenging for control flow hijacking and code-injection attacks. Additionally, Data Space Randomization (DSR)~\cite{codarr} can complicate data-only attacks. While these techniques offer probabilistic guarantees, they significantly raise the difficulty of runtime attacks.

    \circled{4} \textbf{Exploit Dispatch: } To successfully launch sophisticated attacks, the adversary needs to divert the target program to operate on the injected payload. This step is crucial for expressive code-reuse attacks such as ROP, where the attack is initiated by manipulating the stack pointer to execute a sequence of selected gadgets in a predetermined order, with each gadget returning to a specific memory address in the following gadget to implement the desired attack behavior. Control Flow Integrity (CFI)~\cite{original-cfi} and Data Flow Integrity (DFI)~\cite{DFI} are two commonly used defenses to ideally detect and block control flow hijacking and data-only attacks at this stage. These techniques involve implementing and enforcing policies that must be followed during program execution. However, contemporary literature shows that maintaining gap-free policies is inherently challenging, leaving potential exploit opportunities~\cite{cracks,cfi-bending,pop-attack}.

    \circled{5} \textbf{Exploit Execution: } 
    As discussed above, ensuring the complete integrity of a victim program can be challenging. As a result, runtime attestation mechanisms have been proposed as a last line of defense to enable remote verification of code and execution integrity in a trustworthy manner. These mechanisms aim to detect tampering with code or violations of control/data flow.
    In addition, they also provide means to convince a remote party of the execution integrity of the target program during an operation of interest and enable, in some cases, auditing root cause vulnerabilities in case of exploits.
    In addition to measures such as $\textbf{W}\bigoplus\textbf{X}$~\cite{wx} and DEP~\cite{dep} policies, which are deployed to prevent code-injection attacks, remote attestation approaches of code binary~\cite{vrased,android-attest,simple,sailer2004design} are widely regarded as essential for providing remotely verifiable evidence of binary integrity at runtime. At any time during execution, they can be used to attest that the code (including CFI/CFA instrumentation instructions) remains untampered. RA becomes paramount for most-privileged code (and single privilege systems, e.g. bare-metal micro-controllers) where full disablement of runtime code modifications implies the inability to perform remote software updates~\cite{rata}. Control Flow Attestation (CFA)~\cite{c-flat,lofat} and Data Flow Attestation (DFA)~\cite{oat,litehax,dialed} approaches have emerged to specifically detect and audit code-reuse attacks, enabling trustworthy remote verification of control and data flow integrity respectively.


As shown in Figure~\ref{fig:defenses}, attestation mechanisms build atop Roots of Trust (RoTs) as a foundation to provide trustworthy evidence of system/software state that can be remotely verified. For instance, RoTs are utilized in several key aspects of attestation mechanisms. They serve as a foundation for securely measuring system state and/or installed software (RoT for Measurement), securely storing attestation secret keys (RoT for Storage), and/or signing attestation reports (RoT for Reporting). 
Examples include Trusted Platform Modules (TPMs)~\cite{tpm}, DICE~\cite{dice}, hardware extensions in Intel SGX-capable processors~\cite{sgx-explained}, ARM TrustZone-based RoTs~\cite{ARM-TrustZone,TZ_spec}, and various academic proposals such as Keystone~\cite{keystone} and BYOTee~\cite{byotee}, among others.  


\subsection{CFI \& CFA: Definitions \& Threat models}
\label{sec:definitions}

Considering their prominent status as actively researched defenses, the rest of the paper systematically explores CFI and CFA techniques, shedding light on their underlying principles, relationships, trade-offs, and other crucial aspects to provide insights into unclear considerations for adoption in real-world scenarios.


\subsubsection{Control Flow Integrity (CFI)} 


Originally proposed by Abadi et al.~\cite{original-cfi}, CFI is a policy-based mitigation against control flow hijacking attacks, restricting the execution path of a program at runtime based on a pre-computed CFG. In principle, enforcing CFI on a target program involves:

\begin{minimalitemize}
    \item Generation of an over-approximated CFG, denoted $\approx$CFG.

    \item Enforcement of control flow to comply with $\approx$CFG through Reference Monitors (RMs), which are software- or hardware-based runtime checks that verify the target of any indirect branch instruction at runtime.
    

\end{minimalitemize}

$\approx$CFG can be generated statically (as proposed originally~\cite{original-cfi}) or dynamically, as seen in following proposals~\cite{ucfi,pi-cfi}. When $\approx$CFG $\equiv$ CFG, it is generally difficult for an adversary to manipulate control flow and alter a program's intended behavior  without detection by the activated or inserted RMs.
However, statically determining strict CFGs for complex programs remains an open challenge~\cite{where-does-it-go}, leading many practical approaches to over-approximate $\approx$CFG.




\subsubsection{Control Flow Attestation (CFA)}

\begin{figure}[t]
    \centering
    \includegraphics[width=0.8\columnwidth]{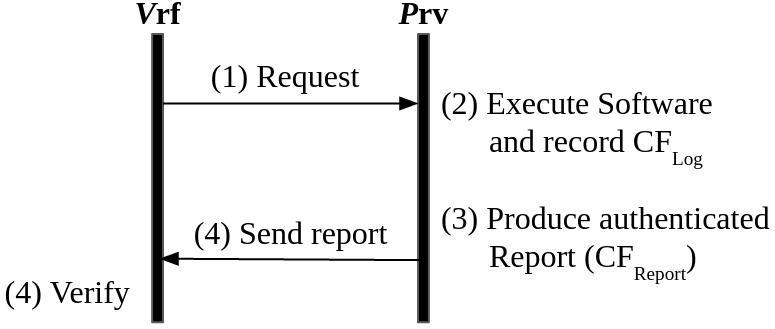}
    \vspace{-0.5em}
    \caption{Typical CFA Interaction}
    \vspace{-1em}
    \label{fig:cfa_prot}
\end{figure}

CFA focuses on producing unforgeable evidence of the control flow path followed by an executable on a prover device (\prv). This evidence allows a remote verifier (\vrf) to assess the trustworthiness of execution and its outcomes. CFA is an (on-demand) challenge-response protocol, as shown in Figure~\ref{fig:cfa_prot}.

A CFA instance starts with \vrf sending a request containing a cryptographic challenge to \prv. Upon receiving the request, \prv must execute the operation requested by \vrf (either specified implicitly or explicitly within the request). During the execution of the requested task, an RoT in \prv must ensure that an authenticated log (\cflog) containing a representation of the control flow path executed during the operation is built. After execution completes, the RoT computes an authenticated integrity measurement (e.g., using a Message Authentication Code (MAC) or signature) over the received challenge, \cflog, and the executed binary to produce a response token (\rep). Finally, \prv transmits \rep to \vrf along with \cflog.
Given the need to securely store the secret used to authenticate \rep even when \prv is potentially compromised, RoT implementations typically involve some form of secure hardware support.

Upon receiving \rep, \vrf can use this evidence to determine if \prv executed the expected software correctly through a valid control flow path. Further, when \rep shows an invalid path, \vrf can analyze the anomalous evidence to determine its cause and potentially remediate it. 

Existing CFA techniques (see Section~\ref{sec:factors}) use either (1) binary instrumentation along with Trusted Execution Environment (TEE) support; or (2) custom hardware modifications to generate \cflog by detecting and saving each branch destination to a dedicated and protected memory region. 
For techniques that use binary instrumentation, a pre-processing phase modifies the binary so that branch instructions are prepended with additional calls to a TEE-protected trusted code. Once called, the trusted code appends to \cflog the current branch destination. In hardware-based techniques, custom hardware interfaces with the CPU to detect branches and record their destinations in protected memory.

\subsubsection{CFI/CFA Coverage} We define the coverage of CFI/CFA proposals in terms of \textit{granularity} and \textit{sensitivity}.

\textbf{Granularity:} The granularity of CFI/CFA mechanisms refers to the detail in which a particular control flow transfer is monitored/checked.
In this work, we categorize the granularity of a specific technique as either \textit{coarse-grained} or \textit{fine-grained}. Since CFI and CFA have different security goals (local detection/prevention vs. providing runtime evidence to a remote party), their granularity pertains to different aspects. 

A \textit{coarse-grained} approach refers to broadly applied checks that are independent of specific control flow transfers within the code. In the case of CFI, this involves techniques applied based on instruction type and agnostic to individual transfers. For instance, the following CFI policies can be classified as \textit{coarse-grained}: enforcing landing pads for calls/returns, checking function type/parameter for indirect calls, and restricting indirect control flow targets within the bounds of a specific sandbox/address space. Since these policies are generally applied to all control flow transfers within a specific scope and do not account for the specific details of each transfer, they are considered \textit{coarse-grained}. In CFA, a scheme is deemed \textit{coarse-grained} if it does not record all control flow transfers within the attested application into \cflog.

A \textit{fine-grained} technique refers to mechanisms that apply a specific check or action for each control flow instruction. In the case of CFI, this entails schemes that verify each indirect target against a unique set of valid locations rather than applying a broader rule based on the instruction type. For instance, enforcement through techniques like shadow stacks, jump-tables, or definition sets determined by data-flow analysis are considered \textit{fine-grained} solutions. A CFA scheme is classified as \textit{fine-grained} if it records all control flow transfers within the attested application.

\textbf{Sensitivity:} Although closely related to granularity, the \textit{sensitivity} of a certain technique describes a different characteristic. It refers to the extent to which execution context is considered for determining the set of valid targets. In this work, we categorize schemes as \textit{insensitive}, \textit{context-sensitive}, or \textit{path-sensitive}. 

Techniques have \textit{insensitive} enforcement if they do not consider the calling context or current execution path when defining the set of valid targets for a particular control flow transfer. As such, the majority of coarse-grained CFI mechanisms are regarded \textit{insensitive} because they employ generic rules, e.g.,  based on the instruction type, ignoring the current execution path or the calling context. 

\textit{Context-sensitive} approaches consider the calling context to determine the set of valid targets. Examples include target bounds being within a particular function/sandbox. Additionally, when a function is called at multiple locations within a second function, a \textit{context-sensitive} approach might determine returns to any call site within the second function as valid. For forward edges, a \textit{context-sensitive} approach allows any valid definition that can reach the function containing the forward edge.

\textit{Path-sensitive} approaches determine the set of valid targets by considering both the calling context and the current executing path. For instance, shadow stacks are regarded as  \textit{path-sensitive} enforcement mechanisms for return addresses because they limit a return to a single call site. 
Furthermore, schemes that employ data flow analysis, such as reaching definitions or points-to analysis, to determine the valid destinations of indirect branches are considered \textit{path-sensitive}.

In CFI, sensitivity affects the local decision on whether a transfer constitutes a violation, whereas in CFA, sensitivity reflects the type of analysis/detection that can be performed by \vrf based on the received evidence.

\subsubsection{CFI/CFA Threat Models \& Assumptions}
\label{subsec:threat}

The security of most CFI techniques depends on the presence of added instrumentation used to enforce CFI checks. In many cases, this is attained via W$\oplus$X permissions for memory accesses, as shown in Figure~\ref{fig:defenses}. While sensible for user-space code, privileged code can typically disable W$\oplus$X enforcement. Therefore, most CFI approaches that target privileged code (e.g., Kernel) rule out code injection/modification from their threat model.

CFA mechanisms require an RoT to implement their attestation functionality, including the acquisition and signing of relevant evidence. The RoT function can also attest the executed binary (and any instrumentation therein) as performed by regular RA. This removes the need for W$\oplus$X enforcement, as long as code is attested in a temporally consistent manner, i.e., code remains the same in the interim between its measurement and execution. This also makes CFA useful to verify privileged code and code that runs on single-privilege Micro Controller Units (MCUs).

Similar to other TEE-based security services, TEE-based CFA (e.g.,~\cite{c-flat,oat,isc-flat}) assumes that any application outside the (hardware-protected) trusted realm of the TEE (e.g., outside the Secure World in TrustZone) can be modified/compromised whereas the RoT implementation within the Secure World is trusted. This is typically supported by a secure boot of the trusted code and implicitly assumes a minimal and vulnerability-free RoT implementation, as vulnerabilities in the RoT can lead to full system compromise~\cite{ret2ns}.
Some CFA methods eliminate the need to trust a software TCB within the TEE by implementing the CFA RoT entirely in hardware~\cite{lofat,litehax,atrium}.

Generally, both CFI and CFA consider the underlying hardware to be trusted, focusing on software-based exploits.
\section{Design Space}
\label{sec:factors}

\begin{figure}[t]
	\centering
	\includegraphics[width=\columnwidth]{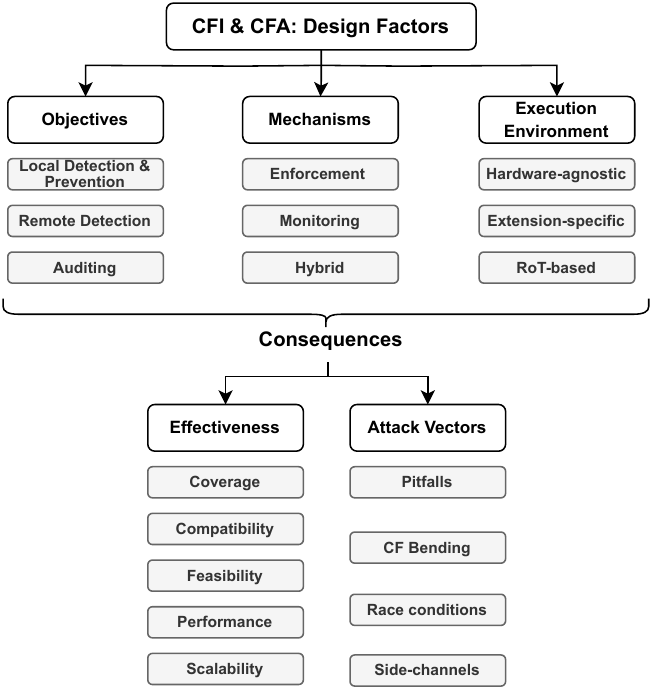}
	\caption{Design Factors of CFI/CFA and Related Consequences}
	\label{fig:factors}
\end{figure}

Figure~\ref{fig:factors} illustrates the distinguishing factors in CFI and CFA, highlighting the consequences of design choices on their effectiveness and susceptibility to attack vectors. Accordingly, Table~\ref{tab:taxonomy} presents a classification of recent work in CFI and CFA, capturing design principles of each mechanism and assessing their trade-offs. Aside from aspects related to security goals (defined in Section~\ref{sec:definitions}), the rest of this section elaborates on design factors. Afterward, Section~\ref{sec:effects} discusses the consequences of these design choices.

\subsection{Different Objectives}


CFI mechanisms primarily focus on \uline{locally detecting} control flow violations during the dispatching stage to \uline{prevent execution of exploited code from continuing}, as depicted in \circled{4} in Figure~\ref{fig:defenses}. In this context, ``prevent'' is not to be confused with the goal of memory safety defenses as mechanisms that aim to remove vulnerabilities (see Section~\ref{sec:scope}).
In other words, CFI does not remove root-cause vulnerabilities. Instead, it impedes certain attack stages, increasing adversaries' difficulty in achieving arbitrary code execution. 

Conversely, CFA is concerned with \uline{providing authentic execution evidence that can be verified and inspected remotely}. In this case, attack detection occurs at a relatively late stage but provides essential insights into attack behavior that can be used to respond to attacks that have evaded prevention measures. This also includes logical bugs (i.e., those not caused by a memory vulnerability) in a program's control flow that CFI would not treat as an exception.
In contrast, CFI does not aim to inform or convince a remote party of execution integrity, handling exceptions and faults locally.

When incorporated atop CFA, runtime auditing~\cite{acfa,traces} aims to reliably deliver evidence to \vrf, even when a compromised \prv attempts not to follow the CFA protocol (see Section~\ref{sec:dots}), refusing to send reports to \vrf in an attempt to hide the exploit behavior.

\subsection{Action Mechanisms}
\label{subsec:mechanisms}


Action mechanisms fall into (i) \uline{enforcement}, (ii) \uline{monitoring} techniques, or (iii) \uline{hybrid} (i.e. a combination thereof) and can be hardware-assisted or implemented in software.  
Early CFI designs relied heavily on \uline{enforcement} through software-based instrumentation (SWI) using generic instructions, so-called Inline Reference Monitors (IRMs)~\cite{bin-cfi,ccfir,llvm-cfi,kcofi,mcfi,ccfi,cfci,ocfi,pi-cfi,lockdown,typearmor,fg-cfi}. 
More recent proposals leverage hardware extensions for specialized CFI instructions as IRMs~\cite{arm-bti,pac-bti-mte,intel-cet,pac-it-up,pac-ret,pacstack,fineibt,pactight,pal}.

A significant limitation in the above-mentioned approaches is the lack of context sensitivity, with transfers checked individually, making these CFI techniques bypassable, as demonstrated in several attacks~\cite{rop-is-dangerous,stitching-the-gadgets,out-of-control,size-does-matter,evaluating-anti-rop,pacman}. This has fueled the development of context-sensitive CFI
~\cite{patharmor,pittypat,ucfi,scfp,os-cfi,cfi-lb,urai,vip-cfi}. Some proposals in this area use advanced points-to-analysis to incorporate path/flow sensitivity to enforce policies effectively. They also leverage commodity hardware features to safeguard the integrity of critical variables that represent the main reference of execution history in such policies~\cite{os-cfi,cfi-lb}.

Hybrid CFI approaches
implement Hardware Reference Monitors (HRMs) using hardware features
to locally save sequences of control flow transfers for asynchronous verification by a separate trusted software module. For instance, PathArmor~\cite{patharmor} leverages the Intel Branch Record (LBR) registers to enable implicit monitoring of execution paths, whereas PittyPAT~\cite{pittypat} and $\mu$CFI~\cite{ucfi} mainly depend on the Intel Processor Tracing (PT) technology~\cite{ipt} to explicitly monitor and verify the execution integrity at runtime. SHERLOC~\cite{sherloc} uses ARM Micro Trace Buffer (MTB) and TrustZone for asynchronous detection of CFI violations.

CFA \uline{monitors} the execution flow, recording transfers to be reported in some form to a \vrf.
C-FLAT~\cite{c-flat} was the first CFA and used software instrumentation to insert IRMs, which redirect each control flow transfer to a secure software routine housed within TrustZone. This routine extends branch destinations into a hash-chain before resuming the attested execution (and performing the branch). 
TinyCFA~\cite{tinycfa} shows an instrumentation-based approach to achieve CFA atop a \textit{Proof of Execution} (PoX) architecture, called APEX~\cite{apex}. Additionally, the work of Papamartzivanos et al.~\cite{multi-level} utilizes Intel PT technology~\cite{ipt} for generating the runtime traces.
LO-FAT~\cite{lofat} and ATRIUM~\cite{atrium} eliminate instrumentation requirements from C-FLAT by implementing custom hardware modules to detect control flow transfers and extend the hash-chain.

While these early approaches produce evidence that minimizes storage/transmission costs (to the size of one hash digest), they result in loss of information, requiring \vrf to use the received hash digest to derive the exact control flow path for inspection. The complexity of this task grows exponentially, leading to the well-known path explosion problem~\cite{aliasing, baldoni2018survey}. 

To ease verification and inspection of CFA evidence, more recent techniques ~\cite{litehax,oat,acfa,scarr,ari,isc-flat} generate \cflog as a lossless trace containing all relevant control flow evidence.
For instance, OAT~\cite{oat}, ARI~\cite{ari}, and TRACES~\cite{traces} leverage TEEs to securely update and store the runtime evidence. LiteHAX~\cite{litehax} and ACFA~\cite{acfa} utilize custom hardware for recording a verbatim trace. 


While CFA techniques are primarily focused on monitoring control flow events passively, recent techniques have proposed \textit{hybrid} action mechanisms that monitor the control flow path while providing some enforcement capabilities. For instance, ISC-FLAT~\cite{isc-flat} creates a TEE-protected dispatcher to ensure that external system interrupts cannot stealthily modify the control flow of the application being attested. CFA+\cite{cfa+} leverages ARMv8.5-A’s landing pad instructions\cite{arm-bti} in combination with selective software instrumentation to enforce a specific CFI policy and enable lightweight monitoring of the execution state, which is maintained in two reserved registers.

\subsection{System Models \& Execution Environments}

CFA and CFI concepts can apply to both high-end systems and low-end embedded devices. Additionally, the requirements of execution environments within the target platform for CFI/CFA can be distinguished as: \uline{hardware-agnostic}, \uline{extension-specific}, and \uline{RoT-based}. 

CFI focuses mainly on the first two types of execution environments, 
and it has primarily been applied in high-end systems. Many earlier CFI approaches are hardware agnostic and utilize SWI to monitor control flow events and locally detect violations. These CFI mechanisms, e.g., LLVM-CFI~\cite{llvm-cfi} and Microsoft Control Flow Guard (MS-CFG)~\cite{ms-cfg}, are easily portable and can cover a variety of high-end targets regardless of the particular CPU type. As they are hardware-agnostic and instrumentation-based, their system models require the presence of a Memory Management Unit (MMU) or an Operating System (OS) to maintain the integrity of the instrumentation. When applied to bare-metal embedded systems that lack MMUs, a Memory Protection Unit (MPU) is used for similar guarantees~\cite{urai,silhouette}.

The portability of CFI in hardware-agnostic environments can come at the price of performance. Hence, several CFI approaches use specific architectural (hardware) extensions in their local environments. Some involve custom hardware extensions specifically designed to support CFI, while others are repurposed from their other goals and integrated as building blocks into CFI. Examples of the former include Intel CET~\cite{intel-cet} and ARM Pointer Authentication~\cite{pac-bti-mte}, along with the body of work built upon them~\cite{fineibt,apple-pac,pac-it-up,pac-ret,pacstack,pactight}. Examples of the latter category include CFI using Intel PT~\cite{ucfi} and ARM Trace Macrocell (TMC)~\cite{kuzhiyelil2020towards}. As extension-specific execution environments are available on both high-end and embedded platforms, CFI proposals using them can also be applied to both classes of target devices.

Most existing CFA approaches target embedded platforms, including those with low-power single-core MCUs (e.g., Atmel AVR ATMega, TI MSP430) with 1-16MHz CPUs, 8-16 bit instruction set architectures (ISAs), used to execute ``bare-metal'' software and commonly implement IoT sensors/actuators and control systems. Additionally, CFA has been proposed for 32-bit embedded platforms with TEE support (e.g., ARM Cortex-M MCUs) that execute real-time applications.
As discussed in Section~\ref{sec:definitions}, CFA necessitates RoT support to generate (and sign) remotely verifiable evidence. RoTs in CFA for embedded devices are implemented via TEEs~\cite{c-flat,oat,isc-flat,traces} or custom hardware changes on \prv~\cite{lofat,litehax,acfa,atrium,tinycfa,rim}. Current CFA proposals aimed at user-space programs~\cite{scarr,recfa,cfa+} either trust the OS (the code integrity of which can be verified using static RA as supported by commodity TPMs) or rely on enclaved execution TEEs~\cite{guarantee}.

\section{Effects \& Consequences}
\label{sec:effects}

This section discusses the effects and consequences of design choices presented in Section~\ref{sec:factors}.

\subsection{Effectiveness}
\label{subsec:effect}

\subsubsection{Coverage} 
In terms of coverage, CFI mechanisms offer varying degrees of protection, ranging from protecting all edges, i.e., all types of indirect control flow altering instructions~\cite{original-cfi,bin-cfi,ccfir,kcofi,mcfi,hcfi,pittypat,griffin}, to partial coverage, targeting either forward-edges~\cite{llvm-cfi,ms-cfg,arm-bti,os-cfi,cfi-lb,vip-cfi,typro,fineibt,pactight} or backward-edges~\cite{hafix,pac-ret,pacstack,urai,silhouette,safe-stack,pac-bti-mte,intel-cet}. Forward-edge~\cite{jop} schemes employ IRMs via SWI~\cite{llvm-cfi,ms-cfg,typro}, hardware-assisted monitoring~\cite{ucfi,pittypat}, landing pads~\cite{arm-bti,fineibt}, and pointer authentication~\cite{pactight,pal,ccfi}. Backward edge~\cite{rop} schemes utilize software-\cite{shadowcallstack,silhouette} or hardware-based~\cite{intel-cet,pac-bti-mte} shadow stacks, and architecture-specific features, such as branch history tables in the x86 processors~\cite{ropecker,kbouncer}, static rewriting, e.g., to form jump tables~\cite{urai}, or pointer authentication~\cite{pac-ret,pacstack}.


Additionally, CFI designs can offer partial protection/coverage in specific scenarios. For example, some designs concentrate on protecting C-like applications~\cite{typro} while leaving out relevant structures specific to C++, such as vtables, and vice-versa~\cite{llvm-cfi}. Conversely, other designs focus on statically linked applications~\cite{original-cfi,bin-cfi,ccfi,ccfir,urai}, overlooking dynamic linking and associated concerns, such as protecting Procedure Linkage Tables (PLT) and Global Offset Tables (GOT)~\cite{gotoverwrite,mcfi}.

The majority of CFI mechanisms (regardless of their coverage) 
are context insensitive~\cite{original-cfi,llvm-cfi,mcfi,ccfi,ccfir,cfi-care,bin-cfi,ms-cfg,fineibt,griffin}. This can introduce gaps that are challenging to detect~\cite{cracks}. Moreover, it limits the ability to detect non-control-data attacks~\cite{non-control-data}.

CFA schemes can enhance coverage and expand (remote) detection capabilities beyond traditional control-flow hijacking attacks. 
As CFA evidence includes the executed control flow path, it (in principle) informs \vrf of any out-of-order execution, including DOP attacks which are oblivious to most CFI. Some approaches~\cite{litehax,oat,dialed} include data inputs within CFA, augmenting produced evidence to also make DDMs observable.
It is important to note, however, that CFA evidence is only truly useful if \vrf can effectively analyze it. This last aspect has been, for the most part, overlooked in the current literature. We revisit this point in Section~\ref{sec:way}.

\begin{table*}[th!]
\centering
\caption{Categorization of CFI and CFA schemes, highlighting their main properties and requirements.}
\label{tab:taxonomy}
\vspace{-0.5em}
\scriptsize
\resizebox{\textwidth}{!}{
\begin{tabular}{|c|c|cccc|cc|c|c|cccc|cccc|}
\hline
\multicolumn{1}{|c|}{\cellcolor[HTML]{EFEFEF}} &
  \multicolumn{1}{c|}{\cellcolor[HTML]{EFEFEF}} &
  \multicolumn{4}{c|}{\cellcolor[HTML]{EFEFEF}\textbf{Device Type/Target}} &
  \multicolumn{2}{c|}{\cellcolor[HTML]{EFEFEF}\textbf{Mechanism}} &
  \multicolumn{1}{c|}{\cellcolor[HTML]{EFEFEF}} &
  \multicolumn{1}{c|}{\cellcolor[HTML]{EFEFEF}} &
  \multicolumn{4}{c|}{\cellcolor[HTML]{EFEFEF}\textbf{Scope}} &
  \multicolumn{4}{c|}{\cellcolor[HTML]{EFEFEF}\textbf{{Overheads}}} \\ 
  \multicolumn{1}{|c|}{\cellcolor[HTML]{EFEFEF}\textbf{Year}} &
  \multicolumn{1}{c|}{\cellcolor[HTML]{EFEFEF}\textbf{Scheme}} &
  
  \multicolumn{1}{c}{\rotatebox{90}{Embedded (bare-metal)\xspace}} &
  \rotatebox{90}{Embedded (OS)} &
   \multicolumn{1}{c}{\rotatebox{90}{High-end (User-space)\xspace}} &
  \multicolumn{1}{c|}{\rotatebox{90}{ High-end (Kernel)}} &
  \multicolumn{1}{c}{\rotatebox{90}{Type}} &
  \multicolumn{1}{c|}{\rotatebox{90}{Strategy}} &
  \multicolumn{1}{c|}{\cellcolor[HTML]{EFEFEF}\textbf{Sensitivity}} &
  \multicolumn{1}{c|}{\cellcolor[HTML]{EFEFEF}\textbf{\textcolor{black}{System Support}}} &
  \rotatebox{90}{Data-Only} &
  \rotatebox{90}{ROP} &
  \rotatebox{90}{JOP} &
  \multicolumn{1}{c|}{\rotatebox{90}{Evidence Expressiveness}} &
  \rotatebox{90}{Runtime} &
  \rotatebox{90}{Code Size} &
  \multicolumn{1}{c}{\rotatebox{90}{Custom Hardware\xspace}} &
  \multicolumn{1}{c|}{\rotatebox{90}{Network}}
  
  \\ \hline

\multicolumn{18}{|c|}{\cellcolor[HTML]{000000}{\color[HTML]{FFFFFF} \textbf{Control Flow Integrity (CFI) Approaches}}} 

\\ \hline

\multicolumn{1}{|l|}{2013} &
  \multicolumn{1}{c|}{bin-CFI~\cite{bin-cfi}} &
  \xmark &
  \xmark &
  \cmark &
  \multicolumn{1}{c|}{\xmark} &
  \multicolumn{1}{c|}{Enforcement} &
  \multicolumn{1}{c|}{SWI} &
  \multicolumn{1}{c|}{\xmark} &
  \multicolumn{1}{c|}{OS/MMU} &
  \xmark &
  \Circle &
  \Circle &
  \multicolumn{1}{c|}{-} &
  \nmarko &
  \nmarkr &
  \xmarkgood &
  \multicolumn{1}{c|}{-} 
  
  \\ \hline
  
\multicolumn{1}{|l|}{2013} &
  \multicolumn{1}{c|}{CCFIR~\cite{ccfir}} &
  \xmark &
  \xmark &
  \cmark &
  \multicolumn{1}{c|}{\xmark} &
  \multicolumn{1}{c|}{Enforcement} &
  \multicolumn{1}{c|}{SWI+R/I} &
  \multicolumn{1}{c|}{\xmark} &
  \multicolumn{1}{c|}{OS/MMU} &
  \xmark &
  \Circle &
  \Circle &
  \multicolumn{1}{c|}{-} &
  \nmarkg &
  \nmarkr &
  \xmarkgood &
  \multicolumn{1}{c|}{-} 
  
  \\ \hline

\multicolumn{1}{|l|}{2014} &
  \multicolumn{1}{c|}{\textbf{LLVM CFI}~\cite{llvm-cfi}} &
  \xmark &
  \xmark &
  \cmark &
  \multicolumn{1}{c|}{\cmark} &
  \multicolumn{1}{c|}{Enforcement} &
  \multicolumn{1}{c|}{SWI} &
  \multicolumn{1}{c|}{\xmark} &
  \multicolumn{1}{c|}{OS/MMU} &
  \xmark &
  \xmark &
  \LEFTcircle &
  \multicolumn{1}{c|}{-} &
  \nmarkg &
  \nmarko &
  \xmarkgood &
  \multicolumn{1}{c|}{-} 
  
  \\ \hline

\multicolumn{1}{|l|}{2014} &
  \multicolumn{1}{c|}{KCoFI~\cite{kcofi}} &
  \xmark &
  \xmark &
  \xmark &
  \multicolumn{1}{c|}{\cmark} &
  \multicolumn{1}{c|}{Enforcement} &
  \multicolumn{1}{c|}{SWI} &
  \multicolumn{1}{c|}{\xmark} &
  \multicolumn{1}{c|}{MMU} &
  \xmark &
  \LEFTcircle &
  \LEFTcircle &
  \multicolumn{1}{c|}{-} &
  \nmarkr &
  \nmarkr &
  \xmarkgood &
  \multicolumn{1}{c|}{-} 
  
  \\ \hline

\multicolumn{1}{|l|}{2014} &
  \multicolumn{1}{c|}{MCFI~\cite{mcfi}} &
  \xmark &
  \xmark &
  \cmark &
  \multicolumn{1}{c|}{\xmark} &
  \multicolumn{1}{c|}{Enforcement} &
  \multicolumn{1}{c|}{SWI} &
  \multicolumn{1}{c|}{CS} &
  \multicolumn{1}{c|}{OS/MMU} &
  \xmark &
  \LEFTcircle &
  \LEFTcircle &
  \multicolumn{1}{c|}{-} &
  \nmarko &
  \nmarko &
  \xmarkgood &
  \multicolumn{1}{c|}{-} 
  
  \\ \hline

\multicolumn{1}{|l|}{2014} &
  \multicolumn{1}{c|}{RockJIT~\cite{rockjit}} &
  \xmark &
  \xmark &
  \cmark &
  \multicolumn{1}{c|}{\xmark} &
  \multicolumn{1}{c|}{Enforcement} &
  \multicolumn{1}{c|}{SWI} &
  \multicolumn{1}{c|}{CS} &
  \multicolumn{1}{c|}{OS/MMU} &
  \xmark &
  \LEFTcircle &
  \LEFTcircle &
  \multicolumn{1}{c|}{-} &
  \nmarkr &
  \nmarko &
  \xmarkgood &
  \multicolumn{1}{c|}{-} 
  
  \\ \hline

\multicolumn{1}{|l|}{2015} &
  \multicolumn{1}{c|}{CCFI~\cite{ccfi}} &
  \xmark &
  \xmark &
  \cmark &
  \multicolumn{1}{c|}{\xmark} &
  \multicolumn{1}{c|}{Enforcement} &
  \multicolumn{1}{c|}{SWI} &
  \multicolumn{1}{c|}{\xmark} &
  \multicolumn{1}{c|}{OS/MMU} &
  \xmark &
  \CIRCLE &
  \CIRCLE &
  \multicolumn{1}{c|}{-} &
  \nmarkr &
  \nmarko &
  \xmarkgood &
  \multicolumn{1}{c|}{-} 
  
  \\ \hline

\multicolumn{1}{|l|}{2015} &
  \multicolumn{1}{c|}{HAFIX~\cite{hafix}} &
  \cmark &
  \xmark &
  \xmark &
  \multicolumn{1}{c|}{\xmark} &
  \multicolumn{1}{c|}{Enforcement} &
  \multicolumn{1}{c|}{ISA} &
  \multicolumn{1}{c|}{Path} &
  \multicolumn{1}{c|}{C-HW} &
  \xmark &
  \CIRCLE &
  \xmark &
  \multicolumn{1}{c|}{-} &
  \nmarkg &
  \nmarkg &
  \nmarko &
  \multicolumn{1}{c|}{-} 
  
  \\ \hline

\multicolumn{1}{|l|}{2015} &
  \multicolumn{1}{c|}{CFCI~\cite{cfci}} &
  \xmark &
  \xmark &
  \cmark &
  \multicolumn{1}{c|}{\xmark} &
  \multicolumn{1}{c|}{Enforcement} &
  \multicolumn{1}{c|}{SWI} &
  \multicolumn{1}{c|}{\xmark} &
  \multicolumn{1}{c|}{OS/MMU} &
  \xmark &
  \Circle &
  \Circle &
  \multicolumn{1}{c|}{-} &
  \nmarkr &
  \nmarko &
  \xmarkgood &
  \multicolumn{1}{c|}{-} 
  
  \\ \hline

\multicolumn{1}{|l|}{2015} &
  \multicolumn{1}{c|}{O-CFI~\cite{ocfi}} &
  \xmark &
  \xmark &
  \cmark &
  \multicolumn{1}{c|}{\xmark} &
  \multicolumn{1}{c|}{Enforcement} &
  \multicolumn{1}{c|}{SWI+R/I} &
  \multicolumn{1}{c|}{\xmark} &
  \multicolumn{1}{c|}{OS/MMU+MPX} &
  \xmark &
  \LEFTcircle &
  \LEFTcircle &
  \multicolumn{1}{c|}{-} &
  \nmarkg &
  \nmarkr &
  \xmarkgood &
  \multicolumn{1}{c|}{-} 
  
  \\ \hline

\multicolumn{1}{|l|}{2015} &
  \multicolumn{1}{c|}{$\pi$CFI~\cite{pi-cfi}} &
  \xmark &
  \xmark &
  \cmark &
  \multicolumn{1}{c|}{\xmark} &
  \multicolumn{1}{c|}{Enforcement} &
  \multicolumn{1}{c|}{SWI} &
  \multicolumn{1}{c|}{\xmark} &
  \multicolumn{1}{c|}{OS/MMU} &
  \xmark &
  \LEFTcircle &
  \LEFTcircle &
  \multicolumn{1}{c|}{-} &
  \nmarkg &
  \nmarkr &
  \xmarkgood &
  \multicolumn{1}{c|}{-} 
  
  \\ \hline

\multicolumn{1}{|l|}{2015} & 
  \multicolumn{1}{c|}{PathArmor~\cite{patharmor}} &
  \xmark &
  \xmark &
  \cmark &
  \multicolumn{1}{c|}{\xmark} &
  \multicolumn{1}{c|}{Hybrid} &
  \multicolumn{1}{c|}{SWI+ISA} &
  \multicolumn{1}{c|}{Path} &
  \multicolumn{1}{c|}{OS/MMU+LBR} &
  \xmark &
  \CIRCLE &
  \CIRCLE &
  \multicolumn{1}{c|}{-} &
  \nmarko &
  \nmarkr &
  \xmarkgood &
  \multicolumn{1}{c|}{-} 
  
  \\ \hline

\multicolumn{1}{|l|}{2015} &
  \multicolumn{1}{c|}{Lockdown~\cite{lockdown}} &
  \xmark &
  \xmark &
  \cmark &
  \multicolumn{1}{c|}{\xmark} &
  \multicolumn{1}{c|}{Enforcement} &
  \multicolumn{1}{c|}{SWI+R/I} &
  \multicolumn{1}{c|}{\xmark} &
  \multicolumn{1}{c|}{OS/MMU} &
  \xmark &
  \CIRCLE &
  \CIRCLE &
  \multicolumn{1}{c|}{-} &
  \nmarkr &
  \nmarkr &
  \xmarkgood &
  \multicolumn{1}{c|}{-} 
  
  \\ \hline

\multicolumn{1}{|l|}{2016} &
  \multicolumn{1}{c|}{TypeArmor~\cite{typearmor}} &
  \xmark &
  \xmark &
  \cmark &
  \multicolumn{1}{c|}{\xmark} &
  \multicolumn{1}{c|}{Enforcement} &
  \multicolumn{1}{c|}{SWI} &
  \multicolumn{1}{c|}{\xmark} &
  \multicolumn{1}{c|}{OS/MMU} &
  \xmark &
  \xmark &
  \LEFTcircle &
  \multicolumn{1}{c|}{-} &
  \nmarkg &
  \nmarkr &
  \xmarkgood &
  \multicolumn{1}{c|}{-} 
  
  \\ \hline

\multicolumn{1}{|l|}{2016} &
  \multicolumn{1}{c|}{FG-CFI~\cite{fg-cfi}} &
  \xmark &
  \xmark &
  \xmark &
  \multicolumn{1}{c|}{\cmark} &
  \multicolumn{1}{c|}{Enforcement} &
  \multicolumn{1}{c|}{SWI} &
  \multicolumn{1}{c|}{\xmark} &
  \multicolumn{1}{c|}{MMU} &
  \xmark &
  \LEFTcircle &
  \LEFTcircle &
  \multicolumn{1}{c|}{-} &
  \nmarkr &
  \nmarkr &
  \xmarkgood &
  \multicolumn{1}{c|}{-} 
  
  \\ \hline

\multicolumn{1}{|l|}{2016} &
  \multicolumn{1}{c|}{HCFI~\cite{hcfi}} &
  \xmark &
  \xmark &
  \cmark &
  \multicolumn{1}{c|}{\xmark} &
  \multicolumn{1}{c|}{Enforcement} &
  \multicolumn{1}{c|}{ISA} &
  \multicolumn{1}{c|}{\xmark} &
  \multicolumn{1}{c|}{OS+C-HW} &
  \xmark &
  \CIRCLE &
  \CIRCLE &
  \multicolumn{1}{c|}{-} &
  \nmarkg &
  \nmarkg &
  \nmarkr &
  \multicolumn{1}{c|}{-} 
  
  \\ \hline

\multicolumn{1}{|l|}{2017} & 
  \multicolumn{1}{c|}{PittyPAT~\cite{pittypat}} &
  \xmark &
  \xmark &
  \cmark &
  \multicolumn{1}{c|}{\xmark} &
  \multicolumn{1}{c|}{Hybrid} &
  \multicolumn{1}{c|}{ISA+HRM} &
  \multicolumn{1}{c|}{Path} &
  \multicolumn{1}{c|}{OS/MMU+PT} &
  \xmark &
  \CIRCLE &
  \CIRCLE &
  \multicolumn{1}{c|}{-} &
  \nmarkr &
  \nmarko &
  \xmarkgood &
  \multicolumn{1}{c|}{-} 
  
  \\ \hline
\multicolumn{1}{|l|}{2017} & 
  \multicolumn{1}{c|}{GRIFFIN~\cite{griffin}} &
  \xmark &
  \xmark &
  \cmark &
  \multicolumn{1}{c|}{\xmark} &
  \multicolumn{1}{c|}{Hybrid} &
  \multicolumn{1}{c|}{ISA+HRM} &
  \multicolumn{1}{c|}{\xmark} &
  \multicolumn{1}{c|}{OS/MMU+PT+TSX} &
  \xmark &
  \CIRCLE &
  \CIRCLE &
  \multicolumn{1}{c|}{-} &
  \nmarkr &
  \nmarko &
  \xmarkgood &
  \multicolumn{1}{c|}{-} 
  
  \\ \hline

\multicolumn{1}{|l|}{2017} &
  \multicolumn{1}{c|}{CFI-CaRE~\cite{cfi-care}} &
  \cmark &
  \xmark &
  \xmark &
  \multicolumn{1}{c|}{\xmark} &
  \multicolumn{1}{c|}{Enforcement} &
  \multicolumn{1}{c|}{SWI+R/I} &
  \multicolumn{1}{c|}{\xmark} &
  \multicolumn{1}{c|}{TZ} &
  \xmark &
  \CIRCLE &
  \LEFTcircle &
  \multicolumn{1}{c|}{-} &
  \nmarkr &
  \nmarkr &
  \xmarkgood &
  \multicolumn{1}{c|}{-} 
  
  \\ \hline

\multicolumn{1}{|l|}{2017} &
  \multicolumn{1}{c|}{\textbf{Intel CET}~\cite{intel-cet}} &
  \xmark &
  \xmark &
  \cmark &
  \multicolumn{1}{c|}{\cmark} &
  \multicolumn{1}{c|}{Enforcement} &
  \multicolumn{1}{c|}{ISA} &
  \multicolumn{1}{c|}{\xmark} &
  \multicolumn{1}{c|}{OS/MMU+CET} &
  \xmark &
  \CIRCLE &
  \Circle &
  \multicolumn{1}{c|}{-} &
  \nmarkg &
  \nmarko &
  \xmarkgood &
  \multicolumn{1}{c|}{-} 
  
  \\ \hline

\multicolumn{1}{|l|}{2018} & 
  \multicolumn{1}{c|}{$\mu$CFI~\cite{ucfi}} &
  \xmark &
  \xmark &
  \cmark &
  \multicolumn{1}{c|}{\xmark} &
  \multicolumn{1}{c|}{Hybrid} &
  \multicolumn{1}{c|}{SWI+ISA} &
  \multicolumn{1}{c|}{\xmark} &
  \multicolumn{1}{c|}{OS/MMU+PT} &
  \xmark &
  \CIRCLE &
  \CIRCLE &
  \multicolumn{1}{c|}{-} &
  \nmarkg &
  \nmarkr &
  \xmarkgood &
  \multicolumn{1}{c|}{-} 
  
  \\ \hline

\multicolumn{1}{|l|}{2018} & 
  \multicolumn{1}{c|}{SCFP~\cite{scfp}} &
  \cmark &
  \xmark &
  \xmark &
  \multicolumn{1}{c|}{\xmark} &
  \multicolumn{1}{c|}{Enforcement} &
  \multicolumn{1}{c|}{SWI+C-HW} &
  \multicolumn{1}{c|}{Path} &
  \multicolumn{1}{c|}{C-HW} &
  \xmark &
  \CIRCLE &
  \LEFTcircle &
  \multicolumn{1}{c|}{-} &
  \nmarko &
  \nmarkr &
  \nmarko &
  \multicolumn{1}{c|}{-} 
  
  \\ \hline

\multicolumn{1}{|l|}{2018} &
  \multicolumn{1}{c|}{\textbf{ARM BTI}~\cite{arm-bti}} &
  \cmark &
  \cmark &
  \cmark &
  \multicolumn{1}{c|}{\cmark} &
  \multicolumn{1}{c|}{Enforcement} &
  \multicolumn{1}{c|}{ISA} &
  \multicolumn{1}{c|}{\xmark} &
  \multicolumn{1}{c|}{BTI} &
  \xmark &
  \xmark &
  \Circle &
  \multicolumn{1}{c|}{-} &
  \nmarkg &
  \nmarko &
  \xmarkgood &
  \multicolumn{1}{c|}{-} 
  
  \\ \hline

\multicolumn{1}{|l|}{2018} &
  \multicolumn{1}{c|}{\textbf{PAC-RET}~\cite{pac-ret}} &
  \cmark &
  \cmark &
  \cmark &
  \multicolumn{1}{c|}{\cmark} &
  \multicolumn{1}{c|}{Enforcement} &
  \multicolumn{1}{c|}{ISA} &
  \multicolumn{1}{c|}{\xmark} &
  \multicolumn{1}{c|}{PA} &
  \xmark &
  \CIRCLE &
  \xmark &
  \multicolumn{1}{c|}{-} &
  \nmarkg &
  \nmarkg &
  \xmarkgood &
  \multicolumn{1}{c|}{-} 
  
  \\ \hline

\multicolumn{1}{|l|}{2019} &
  \multicolumn{1}{c|}{OS-CFI~\cite{os-cfi}} &
  \xmark &
  \xmark &
  \cmark &
  \multicolumn{1}{c|}{\xmark} &
  \multicolumn{1}{c|}{Enforcement} &
  \multicolumn{1}{c|}{SWI+R/I} &
  \multicolumn{1}{c|}{Path} &
  \multicolumn{1}{c|}{OS/MMU+MPX+TSX} &
  \xmark &
  \xmark &
  \CIRCLE &
  \multicolumn{1}{c|}{-} &
  \nmarko &
  \nmarko &
  \xmarkgood &
  \multicolumn{1}{c|}{-} 
  
  \\ \hline

\multicolumn{1}{|l|}{2019} &
  \multicolumn{1}{c|}{CFI-LB~\cite{cfi-lb}} &
  \xmark &
  \xmark &
  \cmark &
  \multicolumn{1}{c|}{\xmark} &
  \multicolumn{1}{c|}{Enforcement} &
  \multicolumn{1}{c|}{SWI+R/I} &
  \multicolumn{1}{c|}{CS} &
  \multicolumn{1}{c|}{OS/MMU+TSX} &
  \xmark &
  \xmark &
  \CIRCLE &
  \multicolumn{1}{c|}{-} &
  \nmarkg &
  \nmarko &
  \xmarkgood &
  \multicolumn{1}{c|}{-} 
  
  \\ \hline

\multicolumn{1}{|l|}{2019} &
  \multicolumn{1}{c|}{PARTS~\cite{pac-it-up}} &
  \xmark &
  \xmark &
  \cmark &
  \multicolumn{1}{c|}{\xmark} &
  \multicolumn{1}{c|}{Enforcement} &
  \multicolumn{1}{c|}{SWI+ISA} &
  \multicolumn{1}{c|}{\xmark} &
  \multicolumn{1}{c|}{OS/MMU+PA} &
  {\bf +} &
  \CIRCLE &
  \CIRCLE &
  \multicolumn{1}{c|}{-} &
  \nmarkr &
  \nmarkr &
  \xmarkgood &
  \multicolumn{1}{c|}{-} 
  
  \\ \hline

\multicolumn{1}{|l|}{2020} &
  \multicolumn{1}{c|}{$\mu$RAI~\cite{urai}} &
  \cmark &
  \xmark &
  \xmark &
  \multicolumn{1}{c|}{\xmark} &
  \multicolumn{1}{c|}{Enforcement} &
  \multicolumn{1}{c|}{SWI+R/I} &
  \multicolumn{1}{c|}{Path} &
  \multicolumn{1}{c|}{MPU} &
  \xmark &
  \CIRCLE &
  \xmark &
  \multicolumn{1}{c|}{-} &
  \nmarkg &
  \nmarkr &
  \xmarkgood &
  \multicolumn{1}{c|}{-} 
  
  \\ \hline

\multicolumn{1}{|l|}{2020} &
  \multicolumn{1}{c|}{Silhouette~\cite{silhouette}} &
  \cmark &
  \xmark &
  \xmark &
  \multicolumn{1}{c|}{\xmark} &
  \multicolumn{1}{c|}{Enforcement} &
  \multicolumn{1}{c|}{SWI+R/I} &
  \multicolumn{1}{c|}{\xmark} &
  \multicolumn{1}{c|}{MPU} &
  \xmark &
  \CIRCLE &
  \xmark &
  \multicolumn{1}{c|}{-} &
  \nmarkg &
  \nmarkg &
  \xmarkgood &
  \multicolumn{1}{c|}{-} 
  
  \\ \hline

\multicolumn{1}{|l|}{2021} &
  \multicolumn{1}{c|}{VIP~\cite{vip-cfi}} &
  \xmark &
  \xmark &
  \cmark &
  \multicolumn{1}{c|}{\xmark} &
  \multicolumn{1}{c|}{Enforcement} &
  \multicolumn{1}{c|}{SWI+R/I} &
  \multicolumn{1}{c|}{Path} &
  \multicolumn{1}{c|}{OS/MMU+MPK} &
  {\bf +} &
  \xmark &
  \CIRCLE &
  \multicolumn{1}{c|}{-} &
  \nmarko &
  \nmarko &
  \xmarkgood &
  \multicolumn{1}{c|}{-} 
  
  \\ \hline

\multicolumn{1}{|l|}{2021} &
  \multicolumn{1}{c|}{PACStack~\cite{pacstack}} &
  \xmark &
  \xmark &
  \cmark &
  \multicolumn{1}{c|}{\xmark} &
  \multicolumn{1}{c|}{Enforcement} &
  \multicolumn{1}{c|}{SWI+ISA} &
  \multicolumn{1}{c|}{\xmark} &
  \multicolumn{1}{c|}{OS/MMU+PA} &
  \xmark &
  \CIRCLE &
  \xmark &
  \multicolumn{1}{c|}{-} &
  \nmarkg &
  \nmarkg &
  \xmarkgood &
  \multicolumn{1}{c|}{-} 
  
  \\ \hline

\multicolumn{1}{|l|}{2022} &
  \multicolumn{1}{c|}{TyPro~\cite{typro}} &
  \xmark &
  \xmark &
  \cmark &
  \multicolumn{1}{c|}{\xmark} &
  \multicolumn{1}{c|}{Enforcement} &
  \multicolumn{1}{c|}{SWI} &
  \multicolumn{1}{c|}{\xmark} &
  \multicolumn{1}{c|}{OS/MMU} &
  \xmark &
  \xmark &
  \CIRCLE &
  \multicolumn{1}{c|}{-} &
  \nmarkg &
  \nmarkg &
  \xmarkgood &
  \multicolumn{1}{c|}{-} 
  
  \\ \hline

\multicolumn{1}{|l|}{2022} &
  \multicolumn{1}{c|}{PAL~\cite{pal}} &
  \xmark &
  \xmark &
  \xmark &
  \multicolumn{1}{c|}{\cmark} &
  \multicolumn{1}{c|}{Enforcement} &
  \multicolumn{1}{c|}{SWI+ISA} &
  \multicolumn{1}{c|}{\xmark} &
  \multicolumn{1}{c|}{PA+MMU} &
  \xmark &
  \CIRCLE &
  \CIRCLE &
  \multicolumn{1}{c|}{-} &
  \nmarkg &
  \nmarkg &
  \xmarkgood &
  \multicolumn{1}{c|}{-} 
  
  \\ \hline

\multicolumn{1}{|l|}{2022} &
  \multicolumn{1}{c|}{PACTight~\cite{pactight}} &
  \xmark &
  \xmark &
  \cmark &
  \multicolumn{1}{c|}{\xmark} &
  \multicolumn{1}{c|}{Enforcement} &
  \multicolumn{1}{c|}{SWI+ISA} &
  \multicolumn{1}{c|}{\xmark} &
  \multicolumn{1}{c|}{OS/MMU+PA} &
  \xmark & 
  \CIRCLE &
  \CIRCLE &
  \multicolumn{1}{c|}{-} &
  \nmarkg &
  \nmarkr &
  \xmarkgood &
  \multicolumn{1}{c|}{-} 
  
  \\ \hline

\multicolumn{1}{|l|}{2023} &
  \multicolumn{1}{c|}{FineIBT~\cite{fineibt}} &
  \xmark &
  \xmark &
  \cmark &
  \multicolumn{1}{c|}{\cmark} &
  \multicolumn{1}{c|}{Enforcement} &
  \multicolumn{1}{c|}{SWI+ISA} &
  \multicolumn{1}{c|}{\xmark} &
  \multicolumn{1}{c|}{CET+MMU} &
  \xmark &
  \xmark &
  \LEFTcircle &
  \multicolumn{1}{c|}{-} &
  \nmarkg &
  \nmarko &
  \xmarkgood &
  \multicolumn{1}{c|}{-} 
  
  \\ \hline

\multicolumn{1}{|l|}{2023} & 
  \multicolumn{1}{c|}{SHERLOC~\cite{sherloc}} &
  \cmark &
  \xmark &
  \xmark &
  \multicolumn{1}{c|}{\xmark} &
  \multicolumn{1}{c|}{Hybrid} &
  \multicolumn{1}{c|}{ISA+HRM} &
  \multicolumn{1}{c|}{Path} &
  \multicolumn{1}{c|}{TZ+MTB+DWT} &
  \xmark &
  \CIRCLE &
  \LEFTcircle &
  \multicolumn{1}{c|}{-} &
  \nmarkr &
  \nmarko &
  \xmarkgood &
  \multicolumn{1}{c|}{-} 
  
  \\ \hline

\multicolumn{1}{|l|}{2023} &
  \multicolumn{1}{c|}{TypeSqueezer~\cite{typesqueezer}} &
  \xmark &
  \xmark &
  \cmark &
  \multicolumn{1}{c|}{\xmark} &
  \multicolumn{1}{c|}{Enforcement} &
  \multicolumn{1}{c|}{SWI} &
  \multicolumn{1}{c|}{Path} &
  \multicolumn{1}{c|}{OS/MMU} &
  \xmark &
  \xmark &
  \CIRCLE &
  \multicolumn{1}{c|}{-} &
  \nmarko &
  \nmarko &
  \xmarkgood &
  \multicolumn{1}{c|}{-} 
  
  \\ \hline

\multicolumn{1}{|l|}{2024} &
  \multicolumn{1}{c|}{HEK-CFI~\cite{hek-cfi}} &
  \xmark &
  \xmark &
  \xmark &
  \multicolumn{1}{c|}{\cmark} &
  \multicolumn{1}{c|}{Enforcement} &
  \multicolumn{1}{c|}{ISA} &
  \multicolumn{1}{c|}{\xmark} &
  \multicolumn{1}{c|}{CET+MMU} &
  \xmark &
  \CIRCLE &
  \LEFTcircle &
  \multicolumn{1}{c|}{-} &
  \nmarkg &
  \nmarkg &
  \xmarkgood &
  \multicolumn{1}{c|}{-} 
  
  \\ \hline

\multicolumn{18}{|c}{\cellcolor[HTML]{000000}{\color[HTML]{FFFFFF} \textbf{Control Flow Attestation (CFA) Approaches}}} 

\\ \hline

\multicolumn{1}{|l|}{2016} &
  \multicolumn{1}{c|}{C-FLAT~\cite{c-flat}} &
  \cmark &
  \xmark &
  \xmark &
  \multicolumn{1}{c|}{\xmark} &
  \multicolumn{1}{c|}{Monitoring} &
  \multicolumn{1}{c|}{SWI} &
  \multicolumn{1}{c|}{\vrf-based} &
  \multicolumn{1}{c|}{TZ} &
  $\square$ &
  \CIRCLE &
  \CIRCLE &
  \multicolumn{1}{c|}{$\triangle$} &
  \nmarkr &
  \nmarkr &
  \xmarkgood &
  \multicolumn{1}{c|}{$\largestar$} 
  
  \\ \hline

\multicolumn{1}{|l|}{2017} &
  \multicolumn{1}{c|}{LO-FAT~\cite{lofat}} &
  \cmark &
  \xmark &
  \xmark &
  \multicolumn{1}{c|}{\xmark} &
  \multicolumn{1}{c|}{Monitoring} &
  \multicolumn{1}{c|}{C-HW} &
  \multicolumn{1}{c|}{\vrf-based} &
  \multicolumn{1}{c|}{C-HW} &
  $\square$ &
  \CIRCLE &
  \CIRCLE &
  \multicolumn{1}{c|}{$\triangle$} &
  \xmarkgood &
  \xmarkgood &
  \nmarkr &
  \multicolumn{1}{c|}{$\largestar$} 
  
  \\ \hline

\multicolumn{1}{|l|}{2017} &
  \multicolumn{1}{c|}{ATRIUM~\cite{atrium}} &
  \cmark &
  \xmark &
  \xmark &
  \multicolumn{1}{c|}{\xmark} &
  \multicolumn{1}{c|}{Monitoring} &
  \multicolumn{1}{c|}{C-HW} &
  \multicolumn{1}{c|}{\vrf-based} &
  \multicolumn{1}{c|}{C-HW} &
  $\square$ &
  \CIRCLE &
  \CIRCLE &
  \multicolumn{1}{c|}{$\triangle$} &
  \xmarkgood &
  \xmarkgood &
  \nmarkr &
  \multicolumn{1}{c|}{$\largestar$} 
  
  \\ \hline

\multicolumn{1}{|l|}{2018} &
  \multicolumn{1}{c|}{LiteHAX~\cite{litehax}} &
  \cmark &
  \xmark &
  \xmark &
  \multicolumn{1}{c|}{\xmark} &
  \multicolumn{1}{c|}{Monitoring} &
  \multicolumn{1}{c|}{C-HW} &
  \multicolumn{1}{c|}{\vrf-based} &
  \multicolumn{1}{c|}{C-HW} &
  $\boxplus$ &
  \CIRCLE &
  \CIRCLE &
  \multicolumn{1}{c|}{$\blacktriangle$} &
  \xmarkgood &
  \xmarkgood &
  \nmarko &
  \multicolumn{1}{c|}{$\filledlargestar$} 
  
  \\ \hline

\multicolumn{1}{|l|}{2019} &
  \multicolumn{1}{c|}{DIAT~\cite{diat}} &
  \cmark &
  \cmark &
  \xmark &
  \multicolumn{1}{c|}{\xmark} &
  \multicolumn{1}{c|}{Monitoring} &
  \multicolumn{1}{c|}{SWI} &
  \multicolumn{1}{c|}{\vrf-based} &
  \multicolumn{1}{c|}{TZ} &
  $\square$ &
  \CIRCLE &
  \CIRCLE &
  \multicolumn{1}{c|}{$\triangle$} &
  \nmarkr &
  \nmarkr &
  \xmarkgood &
  \multicolumn{1}{c|}{$\largestar$} 
  
  \\ \hline

\multicolumn{1}{|l|}{2019} &
  \multicolumn{1}{c|}{ScaRR~\cite{scarr}} &
  \xmark &
  \xmark &
  \cmark &
  \multicolumn{1}{c|}{\xmark} &
  \multicolumn{1}{c|}{Monitoring} &
  \multicolumn{1}{c|}{SWI} &
  \multicolumn{1}{c|}{\vrf-based} &
  \multicolumn{1}{c|}{OS/MMU} &
  $\square$ &
  \CIRCLE &
  \CIRCLE &
  \multicolumn{1}{c|}{$\blacktriangle$} &
  \nmarkr &
  \nmarkr &
  \xmarkgood &
  \multicolumn{1}{c|}{$\filledlargestar$} 
  
  \\ \hline

\multicolumn{1}{|l|}{2019} &
  \multicolumn{1}{c|}{RIM~\cite{rim}} &
  \cmark &
  \xmark &
  \xmark &
  \multicolumn{1}{c|}{\xmark} &
  \multicolumn{1}{c|}{Monitoring} &
  \multicolumn{1}{c|}{C-HW} &
  \multicolumn{1}{c|}{Path} &
  \multicolumn{1}{c|}{C-HW} &
  $\boxplus$ &
  \CIRCLE &
  \CIRCLE &
  \multicolumn{1}{c|}{$\triangle$} &
  \xmarkgood &
  \xmarkgood &
  \textbf{?} &
  \multicolumn{1}{c|}{$\largestar$} 
  
  \\ \hline

\multicolumn{1}{|l|}{2020} &
  \multicolumn{1}{c|}{OAT~\cite{oat}} &
  \cmark &
  \cmark &
  \xmark &
  \multicolumn{1}{c|}{\xmark} &
  \multicolumn{1}{c|}{Monitoring} &
  \multicolumn{1}{c|}{SWI} &
  \multicolumn{1}{c|}{\vrf-based} &
  \multicolumn{1}{c|}{TZ} &
  $\boxplus$ &
  \CIRCLE &
  \CIRCLE &
  \multicolumn{1}{c|}{$\halftriangle$} &
  \nmarko &
  \nmarko &
  \xmarkgood &
  \multicolumn{1}{c|}{$\largestar$} 
  
  \\ \hline

\multicolumn{1}{|l|}{2020} &
  \multicolumn{1}{c|}{LAHEL~\cite{lahel}} &
  \cmark &
  \cmark &
  \xmark &
  \multicolumn{1}{c|}{\xmark} &
  \multicolumn{1}{c|}{Monitoring} &
  \multicolumn{1}{c|}{C-HW} &
  \multicolumn{1}{c|}{\vrf-based} &
  \multicolumn{1}{c|}{C-HW} &
  $\square$ &
  \LEFTcircle &
  \LEFTcircle &
  \multicolumn{1}{c|}{$\halftriangle$} &
  \nmarko &
  \nmarko &
  debug HW &
  \multicolumn{1}{c|}{$\largestar$} 
  
  \\ \hline

\multicolumn{1}{|l|}{2020} &
  \multicolumn{1}{c|}{LAPE~\cite{lape}} &
  \cmark &
  \xmark &
  \xmark &
  \multicolumn{1}{c|}{\xmark} &
  \multicolumn{1}{c|}{Monitoring} &
  \multicolumn{1}{c|}{SWI+R/I} &
  \multicolumn{1}{c|}{\vrf-based} &
  \multicolumn{1}{c|}{MPU} &
  $\square$ &
  \LEFTcircle &
  \LEFTcircle &
  \multicolumn{1}{c|}{$\triangle$} &
  \nmarko &
  \nmarko &
  \xmarkgood &
  \multicolumn{1}{c|}{$\largestar$} 
  
  \\ \hline
  
\multicolumn{1}{|l|}{2021} &
  \multicolumn{1}{c|}{Tiny-CFA~\cite{tinycfa}} &
  \cmark &
  \xmark &
  \xmark &
  \multicolumn{1}{c|}{\xmark} &
  \multicolumn{1}{c|}{Monitoring} &
  \multicolumn{1}{c|}{SWI} &
  \multicolumn{1}{c|}{\vrf-based} &
  \multicolumn{1}{c|}{C-HW} &
  $\square$ &
  \CIRCLE &
  \CIRCLE &
  \multicolumn{1}{c|}{$\blacktriangle$} &
  \nmarkr &
  \nmarkr &
  \xmarkgood &
  \multicolumn{1}{c|}{$\largestar$} 
  
  \\ \hline

\multicolumn{1}{|l|}{2021} &
  \multicolumn{1}{c|}{DIALED~\cite{dialed}} &
  \cmark &
  \xmark &
  \xmark &
  \multicolumn{1}{c|}{\xmark} &
  \multicolumn{1}{c|}{Monitoring} &
  \multicolumn{1}{c|}{SWI} &
  \multicolumn{1}{c|}{\vrf-based} &
  \multicolumn{1}{c|}{C-HW} &
  $\boxplus$ &
  \CIRCLE &
  \CIRCLE &
  \multicolumn{1}{c|}{$\blacktriangle$} &
  \nmarkr &
  \nmarkr &
  \xmarkgood &
  \multicolumn{1}{c|}{$\largestar$} 
  
  \\ \hline

\multicolumn{1}{|l|}{2021} &
  \multicolumn{1}{c|}{ReCFA~\cite{recfa}} &
  \xmark &
  \xmark &
  \cmark &
  \multicolumn{1}{c|}{\xmark} &
  \multicolumn{1}{c|}{Monitoring} &
  \multicolumn{1}{c|}{SWI+R/I} &
  \multicolumn{1}{c|}{\vrf-based} &
  \multicolumn{1}{c|}{OS+MPK} &
  $\square$ &
  \CIRCLE &
  \CIRCLE &
  \multicolumn{1}{c|}{$\blacktriangle$} &
  \nmarkr &
  \nmarkr &
  \xmarkgood &
  \multicolumn{1}{c|}{$\largestar$} 
  
  \\ \hline

\multicolumn{1}{|l|}{2022} &
  \multicolumn{1}{c|}{GuaranTEE~\cite{guarantee}} &
  \xmark &
  \xmark &
  \cmark &
  \multicolumn{1}{c|}{\xmark} &
  \multicolumn{1}{c|}{Monitoring} &
  \multicolumn{1}{c|}{SWI} &
  \multicolumn{1}{c|}{\vrf-based} &
  \multicolumn{1}{c|}{Intel SGX} &
  $\square$ &
  \CIRCLE &
  \CIRCLE &
  \multicolumn{1}{c|}{$\triangle$} &
  \nmarkr &
  \nmarkr &
  \xmarkgood &
  \multicolumn{1}{c|}{$\largestar$} 
  
  \\ \hline

\multicolumn{1}{|l|}{2023} &
  \multicolumn{1}{c|}{ACFA~\cite{acfa}} &
  \cmark &
  \xmark &
  \xmark &
  \multicolumn{1}{c|}{\xmark} &
  \multicolumn{1}{c|}{Monitoring} &
  \multicolumn{1}{c|}{C-HW} &
  \multicolumn{1}{c|}{\vrf-based} &
  \multicolumn{1}{c|}{C-HW} &
  $\square$ &
  \CIRCLE &
  \CIRCLE &
  \multicolumn{1}{c|}{$\blacktriangle$} &
  \xmarkgood &
  \xmarkgood &
  \nmarko &
  \multicolumn{1}{c|}{$\filledlargestar$} 
  
  \\ \hline

\multicolumn{1}{|l|}{2023} &
  \multicolumn{1}{c|}{ARI~\cite{ari}} &
  \cmark &
  \cmark &
  \xmark &
  \multicolumn{1}{c|}{\xmark} &
  \multicolumn{1}{c|}{Monitoring} &
  \multicolumn{1}{c|}{SWI} &
  \multicolumn{1}{c|}{\vrf-based} &
  \multicolumn{1}{c|}{TZ} &
  $\square$ &
  \LEFTcircle &
  \LEFTcircle &
  \multicolumn{1}{c|}{$\halftriangle$} &
  \nmarko &
  \nmarko &
  \xmarkgood &
  \multicolumn{1}{c|}{$\largestar$} 
  
  \\ \hline

\multicolumn{1}{|l|}{2023} &
  \multicolumn{1}{c|}{BLAST~\cite{blast}} &
  \cmark &
  \cmark &
  \xmark &
  \multicolumn{1}{c|}{\xmark} &
  \multicolumn{1}{c|}{Monitoring} &
  \multicolumn{1}{c|}{SWI} &
  \multicolumn{1}{c|}{\vrf-based} &
  \multicolumn{1}{c|}{TZ} &
  $\square$ &
  \CIRCLE &
  \CIRCLE &
  \multicolumn{1}{c|}{$\blacktriangle$} &
  \nmarkr &
  \nmarkr &
  \xmarkgood &
  \multicolumn{1}{c|}{$\largestar$} 
  
  \\ \hline

\multicolumn{1}{|l|}{2023} &
  \multicolumn{1}{c|}{ISC-FLAT~\cite{isc-flat}} &
  \cmark &
  \xmark &
  \xmark &
  \multicolumn{1}{c|}{\xmark} &
  \multicolumn{1}{c|}{Hybrid} &
  \multicolumn{1}{c|}{SWI} &
  \multicolumn{1}{c|}{\vrf-based} &
  \multicolumn{1}{c|}{TZ} &
   $\square$ &
  \CIRCLE &
  \CIRCLE &
  \multicolumn{1}{c|}{$\blacktriangle$} &
  \nmarkr &
  \nmarkr &
  \xmarkgood &
  \multicolumn{1}{c|}{$\largestar$} 

\\ \hline

\multicolumn{1}{|l|}{2024} &
  \multicolumn{1}{c|}{TRACES~\cite{traces}} &
  \cmark &
  \xmark &
  \xmark &
  \multicolumn{1}{c|}{\xmark} &
  \multicolumn{1}{c|}{Monitoring} &
  \multicolumn{1}{c|}{SWI} &
  \multicolumn{1}{c|}{\vrf-based} &
  \multicolumn{1}{c|}{TZ} &
   $\square$ &
  \CIRCLE &
  \CIRCLE &
  \multicolumn{1}{c|}{$\blacktriangle$} &
  \nmarkr &
  \nmarkr &
  \xmarkgood &
  \multicolumn{1}{c|}{$\filledlargestar$} 
  
  \\ \hline

  \multicolumn{1}{|l|}{2024} &
  \multicolumn{1}{c|}{CFA+~\cite{cfa+}} &
  \cmark &
  \cmark &
  \cmark &
  \multicolumn{1}{c|}{\xmark} &
  \multicolumn{1}{c|}{Hybrid} &
  \multicolumn{1}{c|}{SWI+ISA} &
  \multicolumn{1}{c|}{\vrf-based} &
  \multicolumn{1}{c|}{OS/MMU+TPM} &
   $\square$ &
  \CIRCLE &
  \CIRCLE &
  \multicolumn{1}{c|}{$\blacktriangle$} &
  \nmarkg &
  \nmarko &
  \xmarkgood &
  \multicolumn{1}{c|}{$\largestar$} 
  
  \\ \hline

\end{tabular}%
}

\vspace{2pt}

\textbf{Legend:}
\smaller
\cmark \xspace Has this feature, 
\xmark \xspace Lacks this feature, 
\xspace -  Feature is not applicable,
\xspace \textbf{SWI}: Software Instrumentation,
\xspace \textbf{HRM}: Hardware Reference Monitor,
\xspace \textbf{R/I}: Randomization or Isolation,
\xspace \textbf{ISA}: Instruction Set Architecture, 
\xspace \textbf{MMU:} Memory Management Unit,
\xspace \textbf{C-HW}: Custom Hardware,
\xspace \textbf{OS}: Operating System,
\xspace \textbf{CS}: Context Sensitive,
\xspace \textbf{Path}: Context- \& path-sensitive
\xspace \textbf{MPX}: Intel Memory Protection eXtensions
\xspace \textbf{LBR}: Intel Last Branch Record
\xspace \textbf{PT}: Intel Processor Trace
\xspace \textbf{TSX}: Intel Transactional Synchronization Extension
\xspace \textbf{TZ}: ARM TrustZone,
\xspace \textbf{CET}: Intel Control-Flow Enforcement Tech.,
\xspace \textbf{BTI}: ARM Branch Target Identification,
\xspace \textbf{PA}: ARM Pointer Authentication,
\xspace \textbf{MPU}: Memory Protection Unit,
\xspace \textbf{MPK}: Intel Memory Protection keys,
\xspace \textbf{MTB}: ARM Macro Trace Buffer,
\xspace \textbf{DWT}: ARM Data Warchpoint and Trace,
\xspace $+$ Definition-to-use data corruption,
\xspace $\square$ Path deviation DOP,
\xspace $\boxplus$ Definition-to-use data corruption \& path deviation DOP,
\xspace \CIRCLE \xspace Fine-grained,
\xspace \LEFTcircle \xspace Mixed granularity,
\xspace \Circle \xspace Coarse-grained,
\xspace $\triangle$ No path or hashed path, 
\xspace $\halftriangle$ Partial path, 
\xspace $\blacktriangle$ Full path (lossless),
\xspace $\largestar$ All evidence stored and transmitted at once,
\xspace $\filledlargestar$ \xspace Evidence sliced and streamed, 
\xspace \xmarkgood \xspace No overhead, 
\xspace \nmarkg \xspace $<$5\% overhead,
\xspace \nmarko \xspace $<$20\% overhead, 
\xspace \nmarkr \xspace Higher overhead,
\xspace {\bf ?} \xspace Feature required but cost not reported,
\xspace {\bf debug HW} reliance on prototyping/debug features not meant for device deployment.


\end{table*}

\subsubsection{Compatibility}

Compatibility is a fundamental aspect to consider when evaluating the effectiveness of CFI and CFA mechanisms and manifests in various forms. 

\textbf{Binary Support.} Despite the abundance of CFI mechanisms, few operate directly on binary using static binary analysis~\cite{lockdown,typearmor,ccfir,bin-cfi} or specific hardware extensions~\cite{hafix,pittypat,griffin,pt-cfi,ocfi,cfi-care,transparent-rop}. While these have broader applicability, they suffer from false positives~\cite{out-of-control} typically employing ad-hoc approaches to recover CFGs or simply marking all address-taken functions and call-site preceded instructions as legitimate targets for indirect branches~\cite{confirm}. Conversely, CFI based on source code~\cite{llvm-cfi,ucfi,os-cfi,cfi-lb,mcfi,pactight,pal,patharmor} constructs
more accurate CFGs. 
Access to source code typically allows for advanced
static analysis techniques (e.g., points-to analysis as seen in $\mu$CFI~\cite{ucfi} and multi-layer type analysis as seen in MLTA~\cite{where-does-it-go}), increasing coverage and precision. However, source-level schemes do not apply to commercial off-the-shelf (COTS) software, where only binary images are available~\cite{cfi-survey-1}. 

CFA designs typically do not require source code knowledge to generate control flow evidence~\cite{c-flat}, whereas source code knowledge may assist \vrf in analyzing received evidence (see Section~\ref{sec:way}). Hardware-based CFA inherently supports binaries~\cite{lofat,atrium,litehax,acfa} by integrating with the CPU core and detecting branch instructions at runtime. CFA relying on SWI~\cite{c-flat,oat,tinycfa,isc-flat,blast,traces} can instrument control flow transfers in the binary without knowing the source. This is because required instrumentation is used only to log destination addresses rather than determining/enforcing policies in place. Exceptions to this include schemes mixing evidence generation with local integrity checks, e.g.,~\cite{rim,cfa+}. 


\textbf{Modular/Shared Object Support.} A limitation of many CFI mechanisms is the lack of support for external modules or dynamic shared objects (DSO). These mechanisms often rely on global information that may not always be available, making it challenging to implement globally compatible CFI. Abstractly speaking, support for external/shared modules involves (i) integrating multiple modules hardened by CFI separately and (ii) integrating CFI-protected modules with unprotected legacy code. Binary solutions such as CCFIR~\cite{ccfir} attempt to address these issues by allowing more targets than necessary, striking a security-compatibility compromise. Although approaches such as MCFI~\cite{mcfi} and RockJIT~\cite{rockjit} tackle case (i) by independently instrumenting each module and generating new CFGs when modules are linked, recent CFI solutions that offer stronger security guarantees, exemplified by $\mu$CFI~\cite{ucfi} and OS-CFI~\cite{os-cfi}, do not provide modular support. Even contemporary solutions employing hardware features, e.g., PACStack~\cite{pacstack} and PACTight~\cite{pactight}, struggle to address both issues (i) and (ii). 

While not explicitly discussed in prior work, the lack of modular support in CFA can be attributed to (i) most CFA proposals being aimed at simple embedded systems (as seen in Table~\ref{tab:taxonomy}) where applications are statically linked within a single module; and (ii) DSO support would have implications on \vrf evidence analysis, requiring careful consideration.

\textbf{Hardware Dependence.} Hardware-specific features can enhance CFI and CFA. However, they limit a scheme's compatibility to architectures that support them and introduce challenges for legacy systems.
For instance, CFI like PittyPAT~\cite{pittypat}, GRIFFIN~\cite{griffin}, $\mu$CFI~\cite{ucfi}, and PathArmor~\cite{patharmor} utilize Intel PT and LBR to obtain runtime information and compute a smaller set of legitimate targets, striking a balance between accuracy and performance overhead. Similarly,  OS-CFI~\cite{os-cfi}, and CFI-LB~\cite{cfi-lb} leverage Intel TSX (Transactional Synchronization Extensions) and MPX (Memory Protection Extensions) to safeguard instrumented code and metadata against malicious tampering. Approaches such as HCFI~\cite{hcfi} propose custom hardware modifications.

TEE-based CFA schemes demonstrate how instrumentation can be used alongside RoT hardware support (e.g., ARM TrustZone~\cite{ARM-TrustZone}, Intel MPK~\cite{intel-mpk}, or PoX architecture~\cite{apex}) to implement CFA.
Early hardware-based CFA, such as LO-FAT~\cite{lofat} and ATRIUM~\cite{atrium}, add custom branch monitors and hash engines to detect and accumulate control flow transfers as a hash digest. 
LiteHAX~\cite{litehax} opts for more expressive evidence, using dedicated hardware to log and store all control flow transfers, aiming at easing \vrf subsequent analysis. 
ACFA~\cite{acfa} uses custom hardware for branch detection while eliminating the cost of hash engines to make instrumentation-less CFA feasible in budget-constrained micro-controllers. Instead, it incorporates components of a static RA architecture (VRASED~\cite{vrased}) and an active RoT (GAROTA~\cite{garota}). The former is used to authenticate CFA evidence, while the latter is leveraged to ensure reliable delivery of evidence to \vrf (enabling auditing guarantees).

\textbf{Functionality.} Recent evaluations of various CFI defenses have highlighted compatibility issues that can compromise the intended functionality of the target application~\cite{cracks,cfinsight}. Notably, the implementation approach in Lockdown~\cite{lockdown} and OS-CFI~\cite{os-cfi} fails to correctly compile certain applications, e.g., nginx. Moreover, CFI mechanisms such as OS-CFI~\cite{os-cfi} and CFI-LB~\cite{cfi-lb} have been found to generate false positives. 
Additionally, the analysis mechanism of LLVM-CFI~\cite{llvm-cfi} is incompatible with at least one application in the SPEC CPU2006 suite, as reported in~\cite{fineibt}. CFI mechanisms that depend on reserving registers, e.g., VIP~\cite{vip-cfi}, can corrupt functionality when targeting applications with inline assembly that utilize the same registers.

CFA mechanisms that instrument binaries may also encounter compilation failures due to instrumentation issues, as observed in ReCFA with specific benchmarks~\cite{recfa}. As CFA is a newer concept, fewer studies exist on analyzing CFA instrumentation compatibility. At least in principle, issues presented in CFI schemes could also apply to CFA, depending on the instrumentation strategy used. (For instance, similar to VIP-CFI, TinyCFA and TRACES also employ reserved registers.)

\subsubsection{Feasibility} CFI approaches are to a large extent feasible, despite inherent uncertainties around the robustness of policies due to granularity and context sensitivity~(discussed further in  Section~\ref{subsec:attacks}). 
%
As with any attestation mechanism, CFA requires a secure RoT on \prv to maintain and authenticate evidence, as discussed in Section~\ref{subsec:mechanisms}.
It also requires communication with an external \vrf.
%
Naturally, custom hardware features (such as branch monitors) improve feasibility and reduce the cost of CFA.
Being a relatively recent concept, we expect hardware features to support CFA to take longer to reach off-the-shelf devices.

\subsubsection{Performance \& Scalability}
When comparing the scalability of CFI and CFA, it becomes evident that CFI generally encounters fewer or no scalability challenges due to their localized nature. Since the scope of CFI is confined to local decisions based on control flow policies, scalability issues revolve around code size and runtime of the individual applications being protected. A study dedicated to CFI performance can be found in~\cite{cfi-survey-1}.

Performance and scalability in CFA depend on the monitoring strategy. Techniques that utilize SWI incur runtime and code-size increases due to logging~\cite{tinycfa}, context switching into secure environments~\cite{blast}, or both~\cite{c-flat,diat,scarr,oat,lape,ari,traces}. Alternatively, hardware-based CFA mechanisms trade the performance/scalability impact of SWI for additional hardware costs. Early hardware-based approaches impose expensive overheads due to large internal buffers and hardware hash-engines~\cite{atrium,lofat}. Alternative techniques leverage minimal hardware~\cite{litehax,tinycfa} or propose hardware/software co-designs~\cite{acfa} to lower the hardware cost. Regardless of their specificities, existing hardware-based CFA mechanisms are tailored to embedded platforms, leaving scalability in more complex computing platforms as an open challenge.

Unlike CFI, CFA faces further performance challenges in storing and transmitting runtime evidence. Schemes such as ScaRR~\cite{scarr}, ACFA~\cite{acfa}, and TRACES~\cite{traces}, which continuously report evidence to \vrf, may face challenges when attempting to cover multiple active applications on the same \prv. This may impact availability, particularly when network communication is essential (e.g., cloud). In other schemes, \prv may need to store a large \cflog if attested operations are complex. The latter can limit CFA applicability to small and self-contained operations~\cite{oat,tinycfa}. Recent work has investigated dynamically configurable and application-specific optimizations for CFA evidence in MCUs~\cite{speccfa}. Nevertheless, additional research is required to realize CFA (and related optimizations) in higher-end systems.

\subsection{Attack Vectors}
\label{subsec:attacks}
In this section, we explore how gaps or design choices (typically aimed at trading off performance for security) in CFI and CFA can lead to attack vectors.

\subsubsection{Pitfalls}
 Attacks can exploit various well-known pitfalls or limitations, including, but not limited to:

 \textbf{Granularity}: many past attacks have exposed the ineffectiveness of coarse-grained CFI defenses for both forward and backward edges~\cite{rop-is-dangerous,out-of-control,stitching-the-gadgets}. 

 \textbf{Implementation issues}: the implementation of defenses may deviate from their design specifications, leading to a larger number of allowed branch targets than necessary. For example,~\cite{cracks} highlighted implementation mistakes in multiple CFI defenses, including MCFI~\cite{mcfi} and PARTS~\cite{pac-it-up}.

\textbf{Imprecise consideration of language semantics}: Attacks such as COOP~\cite{coop} have affected T-VIP~\cite{tvip} and VTint~\cite{vtint} due to inadequate incorporation of language-specific semantics.

\textbf{Hardware design limitations}: Certain attacks have specifically targeted the hardware design of CFI mechanisms. For instance, the attack on HAFIX~\cite{hafix} highlighted vulnerabilities stemming from hardware limitations~\cite{attack-on-hafix}.

\textbf{Exploitation of Assumptions}: defenses always rely on assumptions within their threat models. Thus, attacks can exploit and falsify these trust assumptions to bypass the defense mechanisms.
For instance, a widespread CFI assumption is W$\oplus$X. The POP attack~\cite{pop-attack} serves as an example where this assumption was violated to bypass FineIBT~\cite{fineibt} defense on the Linux kernel v6.2.8. 

\textbf{Corner Cases}: Certain attacks exploit exceptional cases. For instance, the Control Jujutsu attack~\cite{control-jujutsu} highlighted the limitations of fine-grained CFI defenses with activated shadow stacks in complex code bases like Apache and nginx. Due to coding practices in these code bases, context-insensitive analysis, regardless of its intended robustness, creates over-approximated CFGs that render CFI ineffective. Another example is CHOP~\cite{chop}, which further undermines robust backward edge protection mechanisms, including hardware-based shadow stack implementations~\cite{intel-cet}. It leverages a specific corner case that enables manipulation of the stack unwinding path during exception handling to launch ROP-like attacks, using the unwinder as a confused deputy.


{\bf CFA has not yet been extensively evaluated}: coarser-grained CFA~\cite{ari} (or those based on attesting \prv adherence to locally enforced CFI policies~\cite{rim}) may be subject to the same attack vectors as coarse-grained/context-insensitive CFI, where certain attacks would not appear in the CFA evidence. Yet, CFA that monitors all indirect branches can withstand language semantic issues, enabling detection of attacks such as COOP~\cite{coop}. Additionally, CFA can also provide evidence of logic implementation bugs that lead to unintended paths, in addition to attacks rooted in memory safety vulnerabilities. Naturally, the expressiveness of CFA evidence (i.e., whether it gives \vrf full path evidence or a subset) comes at the price of its (lossless) storage and transmission. Unsurprisingly, implementation deviations (from intended specifications) and falsifiable assumptions would equally affect CFA and CFI.

\subsubsection{Control Flow Bending (CFB)} CFB attacks~\cite{cfi-bending} generalize non-control data attacks targeting CFI schemes relying on statically generated CFGs.
While many CFI attacks target weaker or sub-optimal implementations~\cite{out-of-control,stitching-the-gadgets,rop-is-dangerous}, 
CFB focuses on bypassing the most restrictive (or optimal) static CFI policies.
CFB creates malicious (Turing-complete) paths that exist on the most strict CFG for a given program by exploiting specific functions, called \textit{dispatchers}, which have the capability to modify their own return addresses. In other words, CFB can arbitrarily modify (bend) a program's behavior/path while staying within the confines of the imposed security policy.


This highlights that even fine-grained CFI can be bypassed if dynamic backward protection is not implemented (e.g., via a secure shadow stack).
To mitigate CFB, certain CFI proposals incorporate dynamic analysis~\cite{pi-cfi} or leverage hardware features that provide runtime information on execution status~\cite{ucfi}. Additionally, context-sensitive CFI schemes have the potential to reduce the impact of CFB by maintaining an execution history and validating the execution of return instructions accordingly~\cite{os-cfi, cfi-lb}.

Most CFA approaches log all dynamically defined branch targets within their execution scope. 
Thus, CFB path deviations appear in generated evidence, making CFB attacks apparent to \vrf. 
That said, (similar to cases discussed above) the effectiveness \vrf in detecting CFB based on CFA evidence remains to be concretely evaluated.

\subsubsection{Race Conditions} 

Many CFI methods overlook thread safety in multi-threaded applications. This can leave them vulnerable to Time-Of-Check-to-Time-Of-Use (TOCTOU) attacks. Software-based approaches such as LLVM-CFI~\cite{llvm-cfi} face challenges in ensuring thread safety, especially in the presence of blind compiler optimizations that can inadvertently expose sensitive variables used for security checks. This can create race conditions that enable TOCTOU attacks~\cite{losing-control}.

Additionally, WarpAttack~\cite{warpattack} revealed that compiler optimizations could introduce double-fetch vulnerabilities, resulting in concurrency issues and TOCTOU, even with a strict static CFI policy that includes both forward and backward-edge protections. WarpAttack bypassed several CFI defenses, including LLVM-CFI~\cite{llvm-cfi}, Lockdown~\cite{lockdown}, and MS-CFG~\cite{ms-cfg}. To mitigate race conditions, contemporary CFI mechanisms rely on hardware support. For example, OS-CFI~\cite{os-cfi} and CFI-LB~\cite{cfi-lb} utilize Intel TSX to safeguard intermediate values.

In CFA (and more broadly RA), resistance against TOCTOU attacks and race conditions often refers to achieving temporal consistency between when the executable binary is measured and when it is executed~\cite{temporal-consistency,toctou-ra,hristozov2022toctou,tinycfa}.
Aside from modifications to code, the integrity of CFA evidence can be compromised by external interrupts that may stealthily modify the control flow path or the execution state, as shown and mitigated by ISC-FLAT~\cite{isc-flat}.

\subsubsection{Side channels} The emergence of microarchitectural attacks can affect CFI and CFA. While these defenses focus on memory corruption attacks, certain variants of Spectre~\cite{spectre} can affect them. For instance, Spectre v1 exploits misspeculation following a bounds-check prior to an array access, while Spectre v2 exploits misprediction of the target of an indirect call or jump. Both utilize a Flush+Reload channel~\cite{flush+reload} to leak data. Research has demonstrated that Spectre v1-like attacks can bypass software-based CFI defenses, such as LLVM-CFI~\cite{llvm-cfi}, even in the presence of all default mitigations~\cite{bypassing-llvmcfi-spectre}. While specialized mitigations, such as S\textsc{pec}CFI~\cite{speccfi} and MicroCFI~\cite{microcfi}, were proposed, Spectre v2 remains severe and yet to be fully mitigated.

Although contemporary CFI defenses, such as Intel CET~\cite{intel-cet}, consider a post-Spectre threat model and are designed with built-in protection against Spectre v2~\cite{sepctre-analysis-cet}, recent attacks, such as InSpectre Gadget~\cite{inspectre}, have uncovered new types of exploitable gadgets that can successfully mount Spectre v2 attacks, even if the CET's Indirect Branch Tracking (IBT) feature or its recent fine-grained counterpart, FineIBT~\cite{fineibt}, are active. PACMAN~\cite{pacman} stands out as another recent attack that exploits speculative execution along with memory corruption to bypass ARM Pointer Authentication on Apple M1 SoCs.

Similar to CFI, CFA leveraging architectural components vulnerable to side channels could be equally vulnerable.
On the other hand, several secret dependency-related time side channels~\cite{timing} (that exploit software implementation bugs, rather than micro-architectural bugs) depend on differences in the target program's control flow path, opening opportunities for exploit identification based on \cflog analysis. To our knowledge, the latter remains unexplored in prior work.

\section{Takeaways and Paths Forward}
\label{sec:dots}




We conclude this paper synthesizing insights from discussions presented in Section~\ref{sec:factors}, Section~\ref{sec:effects}, and 
Table~\ref{tab:taxonomy}. Based on these insights, we revisit questions {\bf [Q1-Q4]} from Section~\ref{sec:introduction}.



\subsection{Takeways}

Considering question {\bf [Q1]} posed in Section~\ref{sec:introduction}, this systematization presents several differences between CFI and CFA.
The effectiveness of CFI mechanisms is intrinsically tied to the comprehensiveness and accuracy of a (statically-defined or dynamic) policy enforced locally. Most CFA techniques are policy-agnostic, passively monitoring execution to generate authenticated control flow reports.
Contrary to CFI, CFA concerns convincing a remote party of trustworthy execution behavior, serving as a runtime analog to static attestation methods that prove the integrity of booted/loaded code. Thus, CFA reports are transmitted to a remote \vrf for analysis. These observations lead us to Takeaway 1.

Regarding \textbf{[Q2]}, we first examine CFA/CFI assumptions. Many CFI schemes assume the ability to apply W$\oplus$X on memory to preserve the integrity of IRMs and avoid code injection. In Table~\ref{tab:taxonomy}, this is apparent from user-space CFI schemes frequently relying on OS/MMU system support to enforce the W$\oplus$X policy. While CFA mechanisms need not impose W$\oplus$X, they must rely on an attestation RoT in \prv to attest that reported runtime evidence is authentic (this includes code integrity and instrumentation, when applicable). Furthermore, unlike CFI, CFA requires network connectivity between \vrf and \prv. Despite these differences, we also observe that state-of-the-art techniques for CFI and CFA intersect in their mechanisms for monitoring control flow events. 
For instance, many schemes utilize IRMs via SWI as a mechanism while relying on hardware (whether commodity or custom) to protect or support their instrumentation, as shown in Table~\ref{tab:taxonomy}. Table~\ref{tab:taxonomy} also shows that both CFI and CFA can optimize SWI using specific ISA extensions. This is summarized in Takeaway 2. 

\begin{figure}[t]
  \begin{tcolorbox}[colback=Gray2,colframe=darkcerulean,title=\textup{\textbf{Takeaway 1: CFI and CFA have different goals}}]
  CFI focuses on local detection of control-flow violations, whereas CFA provides remote evidence of execution behavior irrespective of underlying policy enforcement.
  \end{tcolorbox}
\end{figure}


\begin{figure}[h!]
  \begin{tcolorbox}[colback=Gray2,colframe=darkcerulean,title=\textup{\textbf{Takeaway 2: Design intersections \& differences}}]
  Although CFI and CFA schemes share many commonalities in their strategies (as apparent in the \texttt{Mechanism} column of Table~\ref{tab:taxonomy}), they also have distinct requirements for their system models, e.g., as seen in the \texttt{System Support} and \texttt{Network Overhead} columns of Table~\ref{tab:taxonomy}. 
  \end{tcolorbox}
\end{figure}

A common misconception/over-simplification that relates to {\bf [Q3]} is that CFA's entire purpose is to enable CFI checks to be outsourced to a resource-rich \vrf, avoiding CFI costs on \prv. As extensively discussed in this systematization, CFA goals go beyond outsourcing CFI checks. 
As evidence of that, recent CFA methods have evolved to generate expressive (often lossless) control flow path evidence, as opposed to proving adherence to a locally enforced CFI policy (see \texttt{Evidence Expressiveness} column, in Table~\ref{tab:taxonomy}).
This is subsumed by Takeaway 3.


\begin{figure}[t]
  \begin{tcolorbox}[colback=Gray2,colframe=darkcerulean,title=\textup{\textbf{Takeaway 3: CFA goes beyond outsourced CFI}}]
   While CFI is clearly the best choice for local detection of runtime attacks, CFA enables remote (and offline) control flow path analysis, giving remote visibility to complex path deviations (e.g., Control Flow Bending) that would often be oblivious to most CFI -- see \texttt{Scope} column in Table~\ref{tab:taxonomy}. CFA evidence also makes logic control path bugs (other than memory corruption) observable. Finally, it facilitates auditing and root cause analysis if the evidence is reliably delivered to \vrf. On the other hand, remote observability in CFA comes at the cost of supporting communication and securely implementing an attestation RoT. 
  \end{tcolorbox}
  \vspace{-1em}
\end{figure}

Regarding {\bf [Q4]}, given their distinct security goals, the coexistence of CFI and CFA on the same platform could be possible if the performance overhead is acceptable in the target domain. We believe the exploration of approaches that combine the strengths of CFI and CFA to be an intriguing avenue for further research. A potential hybrid design might include CFI building blocks that can be elegantly incorporated into CFA reports. Considering that many state-of-the-art CFI offers fine-grained local ROP detection with low overheads (as seen in \texttt{Scope} and \texttt{Overheads} columns in Table~\ref{tab:taxonomy}), a hybrid approach might implement CFI techniques for local ROP detection while utilizing CFA techniques for generating expressive evidence of path deviations due to JOP, and/or logic control bugs. Yet, CFI/CFA integration is non-trivial, as differences in designs and system assumptions should be considered and can contribute to overheads.
As a first step in this direction, CFA+\cite{cfa+} recently proposed a mechanism that combines CFI and CFA to locally enforce specific targets for certain control flow transfers, relying on ARMv8.5-A landing pad instructions\cite{arm-bti} while leveraging minimal instrumentation to record path information in reserved registers.
These observations are summarized in Takeaway 4.


\begin{figure}[h!]
  \begin{tcolorbox}[colback=Gray2,colframe=darkcerulean,title=\textup{\textbf{Takeaway 4: Coexistence merits investigation}}]
   Given the trade-offs between CFI and CFA, a hybrid approach could offer both local responses to simpler runtime attacks and remote visibility to complex attacks and their root causes. On the other hand, overheads of both approaches on the same platform could challenge practical adoption. 
  \end{tcolorbox}
  \vspace{-1em}
\end{figure}

\subsection{Paths Foward}
\label{sec:way}


\textbf{Demand for Stronger Threat Models}. Currently considered threat models (in both CFI and CFA) can be limited in scope or may not adequately address the challenges posed by sophisticated adversaries (e.g., those capable of launching side-channel attacks). Next-generation mechanisms could consider stronger threat models to encompass new attack vectors that can lead to control-flow violations.

\textbf{CFA Support for Complex Software}. The current landscape of CFA mechanisms primarily focuses on addressing the needs of simple, specialized, bare-metal embedded software (see column \texttt{Device Type/Target} in Table~\ref{tab:taxonomy}). This limited scope poses challenges when it comes to applying these mechanisms to complex software scenarios with wider attack surfaces. To overcome this limitation, it is crucial to develop CFA mechanisms specifically tailored for complex software.

\textbf{CFA Evidence Verification \& Practicality.} The majority of CFA literature focuses on \prv, assuming a \vrf can interpret received evidence to detect attacks and identify root causes as long as the evidence is sufficiently expressive. Alas, there is a significant lack of concrete \vrf instances to substantiate postulated evidence analysis capabilities. Most of the CFA literature either leaves \vrf implementation as future work or implements simple remote checks based on received evidence, e.g.,  adherence to a CFG or emulated shadow stack (both of these could also be done locally by several CFI methods and still face the complex challenge of validating non-deterministic forward-edges).
Only two studies have explored alternative approaches to simple remote checks. ZEKRA~\cite{zekra} suggests generating a zero-knowledge proof of CFG adherence for an untrusted \vrf, while RAGE~\cite{rage} proposes training a Graph Neural Network (GNN) on previous runtime evidence for path verification. Yet, thus far, no prior work has concretely demonstrated CFA's postulated benefits in uncovering complex attacks (and their root causes) based on remotely analyzed evidence.
Additionally, striking a balance between evidence expressiveness and overhead poses a challenge in achieving full-fledged CFA. Hashed paths compromise detailed runtime evidence in exchange for reduced storage and transmission costs. However, lossless path representations (and associated transmission to \vrf) remain costly. As the complexity of the applications increases, the importance of expressiveness/cost trade-offs becomes more pronounced. Within this realm, promising avenues for future work include the development of mechanisms to reduce evidence storage and transmission costs while maintaining relevance and expressiveness. In this direction, recent work in SpecCFA~\cite{speccfa} proposes architectural support for \vrf-defined application-specific optimizations based on likely control flow sub-paths, enabling reduced storage/transmission costs without compromising evidence expressiveness. 

\textbf{CFI in Real-Time Systems and Other Niche.} 
%
%
Most CFI proposals in Table~\ref{tab:taxonomy} are not well-suited for real-time systems where strict timing requirements and execution integrity must be simultaneously maintained. 
Several recent CFI proposals focus on this gap~\cite{insectacide,fastcfi,recfish,kadar,ecfi}.
InsectACIDE~\cite{insectacide} uses architectural support (ARM TrustZone and MTB) to record control flow events without adding intra-task delays. During idle periods in between the execution of tasks, InsectACIDE uses the recorded information to perform security checks and locally detect control flow violations. FastCFI~\cite{fastcfi} also uses ARM features for tracing control flow transfers but relies on Field Programmable Gate Arrays (FPGA) to store and traverse a CFG according to these transfers. ECFI~\cite{ecfi} proposes a mechanism for real-time Programmable Logic Controllers (PLCs) in which indirect branches are instrumented, and a PLC's OS can schedule CFI checks. In ECFI, CFI checks are assigned a lower priority so that PLC tasks maintain their Worst-Case Execution Time (WCET). RECFISH~\cite{recfish} proposes CFI for ARM MCUs executing both bare-metal software or applications atop FreeRTOS. RECFISH utilizes SWI to insert trampolines to functionality enforcing indirect branch destinations according to function label sets or a shadow stack. When applied to FreeRTOS, RECFISH also saves the task's state to the shadow stack to ensure that CFI-critical data cannot be overwritten during a context switch.

Another challenge involves rethinking the conventional approach of terminating an exploited application upon detecting a CFI violation, especially in domains such as autonomous systems. Abruptly terminating an application can introduce system instability or disruptions, posing risks to critical operations. Current proposals for CFI in real-time systems generally focus on minimizing WCET while supporting local detection rather than delving into post-detection recovery strategies~\cite{fastcfi,insectacide,recfish}. One alternative approach in ECFI~\cite{ecfi} makes killing the violating process configurable via an optional flag and, by default, stores a log file of the violation details. Alternative strategies could be explored to recover from CFI violations and ensure system safety without unintended consequences. A promising direction to address this challenge is to design CFI schemes accommodating multi-variant execution that allows the containment of exploited applications while enabling the continuation of critical tasks. 
We note that devising CFI that tackles both aforementioned challenges is non-trivial and requires more careful consideration. This includes accounting for factors such as portability, adaptability, and scalability.




\section*{Acknowledgements}

The authors sincerely thank the anonymous shepherd and the anonymous reviewers for their guidance and constructive criticism.

\IEEEtriggeratref{175}

\bibliographystyle{IEEEtran}
\bibliography{small-refs}


\newpage 



\section*{META-REVIEW}

The following meta-review was prepared by the program committee for the 2025 IEEE Symposium on Security and Privacy (S\&P) as a part of the review process as detailed in the call for papers.

\subsection*{Summary}
This paper provides an SoK on CFI and CFA. The work is motivated by the increasing threat of memory corruption vulnerabilities and the wide scope of defenses that have been proposed to mitigate them. The defenses fall under two categories, CFI and CFA, but are riddled with various goals, assumptions, and implementation weaknesses. This paper disentangles the literature to provide a unified view of the two approaches, their strengths and weaknesses, and promising directions for future research.

\subsection*{Scientific Contributions}
\begin{minimalitemize}
    \item Independent Confirmation of Important Results with Limited Prior Research
    \item Provides a Valuable Step Forward in Established Field
\end{minimalitemize}

\subsection*{Reasons for Acceptance}
\begin{minimalenum}
    \item The community is in need of a comparison and deep understanding of the difference and gap of CFI and CFA.
    \item The paper does a great job at comparing and contrasting CFI and CFA and positioning them within the broader scope of runtime defenses.
    \item It also does a good job of highlighting promising avenues for future work, including problem domains, high-level abstractions, and low-level techniques.
\end{minimalenum}




\end{document}